\documentstyle[aps,preprint,tighten,eqsecnum,epsf]{revtex}

\newcommand{\bc}{\begin{center}}
\newcommand{\ec}{\end{center}}
\newcommand{\nin}{\noindent}
\newcommand{\be}{\begin{equation}}
\newcommand{\ee}{\end{equation}}
\newcommand{\ba}{\begin{array}}
\newcommand{\ea}{\end{array}}

\newcommand{\tw}{\tau_{\scriptscriptstyle W}}

\newcommand{\HT}{\tilde{H}}

\begin{document}
\draft
%\preprint{}
%%%%%%%%%%%%%%%%%%%%%%%%%%%%%%%%%%%%%%%%%%%%%%%%%%%%%%%%%%%%%%%%%%%%%%%%%
\title{Parametric statistics of the scattering matrix: From metallic to
insulating  quasi-unidimensional disordered systems}

\author{E. R. Mucciolo $^{(1)}$, R. A. Jalabert $^{(2,3)}$, and
        J.-L. Pichard $^{(2,4)}$}

\address{$^{(1)}$ Departamento de F\a'{\i}sica, Pontif\a'{\i}cia Universidade
        Cat\'olica, C.P. 38071, 22452-970 Rio de Janeiro, RJ, Brazil}

\address{$^{(2)}$ Institute for Theoretical Physics, University of California,
        Santa Barbara, CA 93106-4030}

\address{$^{(3)}$ Universit\'e Louis Pasteur, IPCMS-GEMME, 23 rue du Loess,
        67037 Strasbourg Cedex, France}

\address{$^{(4)}$ Service de Physique de l'Etat Condens\'e,
        CEA Saclay, 91191 Gif-sur-Yvette, France}

\date{\today}

\maketitle

\begin{abstract}
We investigate the statistical properties of the scattering matrix $S$
describing the electron transport through quasi-one dimensional
disordered systems. For weak disorder (metallic regime), the energy
dependence of the phase shifts of $S$ is found to yield the same
universal parametric correlations as those characterizing chaotic
Hamiltonian eigenvalues driven by an external parameter. This is
analyzed within a Brownian motion model for $S$, which is directly
related to the distribution of the Wigner-Smith time delay matrix.
For large disorder (localized regime), transport is dominated by
resonant tunneling and the universal behavior disappears. A model
based on a simplified description of the localized wave functions
qualitatively explains our numerical results. In the insulator, the
parametric correlation of the phase shift velocities follows the
energy-dependent autocorrelator of the Wigner time. The Wigner time
and the conductance are correlated in the metal and in the insulator.

\end{abstract}


%%%%%%%%%%%%%%%%%%%%%%%%%%%%%%%%%%%%%%%%%%%%%%%%%%%%%%%%%%%%%%%%%%%%%%%%%
\section{Introduction}
\label{sec:intro}

%------------------------------------------------------------------------
\subsection {Preface}

Electron transport through quasi-one dimensional disordered systems is
characterized by the scattering matrix $S$, relating the amplitudes of
incoming and outgoing waves. The scattering matrix contains
information about not only quantities as the conductance and the
characteristic dwelling times for electrons moving through the
disordered region, but also on the energy levels of the system. The
universal statistical properties found in the energy spectrum of
disordered and chaotic systems have been related to the distribution
of eigenvalues of random matrix Hamiltonians \cite{Efe,AS,BGS}. On the
other hand, the universal conductance fluctuations of metallic systems
have been understood in terms of a statistical description of the
transfer matrix \cite{St,Been2}. To understand the relation between
these two universalities, it is useful to study the statistical
properties of the scattering matrix
\cite{VerbWeidZirn,IWZ,meba,ropibe2,Weigr,JP}.

Due to current conservation, the scattering matrix is unitary and its
eigenvalues are represented by $2N$ phase shifts $\{\theta_l\}$. ($N$
is the number of transverse channels in each asymptotic region.)
Assuming that all matrices $S$ with a given symmetry are equally
probable, we obtain Dyson's circular ensembles \cite{Dy,Mehtabook}
(named COE and CUE for orthogonal and unitary symmetry classes,
respectively), where the phase shift statistics follows universal
laws. The isotropy hypothesis of Dyson's ensembles applies only to
ballistic chaotic cavities with no direct channels \cite{BlSm,meba},
where the electronic motion is essentially zero-dimensional after a
(short) time of flight. The conductance in this systems is always of
the order of $N/2$ since reflection and transmission are equally
likely. On the other hand, in disordered systems this isotropy
hypothesis is not satisfied any longer because transmission is much
less probable than reflection. For quasi-one-dimensional (quasi-1d)
metallic samples the conductance is of the order of $Nl/L$, where $l$
is the elastic mean-free-path, $L$ the length of the disordered
region, and $l\!\ll\!L$. In the localized regime, when the
localization length $\xi\sim Nl$ is much smaller than the sample
length $L$, the typical conductance scales as
$\exp{(-2L/\xi)}$. However, the failure of Dyson's hypothesis for the
mean values of $S$ (which is linked to the average conductance) does
not prevent the fluctuations of $S$ for weakly-disordered quasi-1d
samples to be (approximately) described with the universal behavior of
the circular ensembles, as shown in Ref.~\cite{JP}. This good
agreement gets poorer if the disorder is increased, as well as when
the system does not have a quasi-1d geometry, or when it enters the
localized regime. In the first two cases the mean values of the
transmission and reflection amplitudes become strongly dependent on
the channel index and the eigenphase distribution is highly
anisotropic. When localization is achieved by keeping the disorder
weak and increasing the length of the sample beyond the localization
length, the scattering matrix was shown to decouple in two
statistically independent (almost unitary) reflection matrices
\cite{JP}. This picture is particularly clear if one uses a
semiclassical approach where most of the classical trajectories return
to the region of departure instead of traversing the disordered
sample.

A few years ago, another type of universality in the spectrum of
chaotic and disordered systems was discovered by Szafer, Altshuler,
and Simons \cite{Aaron,Ben}. It concerns the adiabatic response to an
external perturbation (magnetic field, shape of the confining
potential, etc). The correlator of the derivatives of the
eigenenergies at two different values of the external parameter has a
universal functional form once a proper rescaling is carried
out. Latter studies have found these universal parametric correlations
to hold in a variety of other systems, including interacting
one-dimensional models \cite{Michael}, $\vec{k}$-dependent band
structures of semiconductors \cite{Mucc}, Rydberg atoms in magnetic
fields \cite{Kleppner}, etc. Considering the $2N$ eigenphases of $S$
as a function of the electron energy, the natural question arises as
to whether the analogous parametric correlations will have the
universal form found for eigenenergies. In the weakly-disordered
quasi-1d case, where the fluctuations of the eigenphases follow the
circular ensemble predictions and $N\gg1$, one could anticipate that
the parametric correlations do exhibit the universal form of Szafer
{\it et al.}, and this is confirmed here by numerical simulations and
analytical arguments. More interesting is the question of what happens
to these correlations as we go into the localized regime, and whether
or not they can be described by a simple model as that of the
decoupling of the eigenphases.

Directly related to the parametric correlations of the eigenphases,
there is the question of the statistical distribution of the
traversing-times in the disordered region. This distribution and the
related correlation functions have recently begun to be addressed in
Refs. \cite{LSSS,FS}, because the Wigner time appears in a variety of
physical phenomena, ranging from the capacitance of mesoscopic quantum
dots \cite{Gopar} to nuclear resonances \cite{Smith}. Since the Wigner
time is obtained from the scattering matrix and its energy derivative,
the statistical properties of $S(E)$ determine the correlations of the
traversal time. We aim in this paper to study the relationship between
the parametric correlations and the distributions of the traversal
times, and their connection to the statistical properties of the
corresponding scattering matrix for metallic and localized quasi-1d
disordered systems.

In the remaining of this section we introduce the basic notation
concerning the scattering matrix and present the substantially
different behavior of the eigenphases, conductance, and Wigner time
(as functions of the Fermi energy) between the metallic and localized
cases. In Sec.~\ref{sec:metallic} we study the parametric correlations
of weakly-disordered quasi-1d systems. We justify their universal
character in the framework of a Brownian-motion model for unitary
matrices. We then study the energy correlations of the conductance and
the Wigner time and the cross-correlation between these two
quantities, making contact with existing calculations in the metallic
regime. In Sec.~\ref{sec:localized} we undertake similar studies for
the localized regime. The non-universal parametric correlations of
eigenphases and Wigner times can be accounted for, at the
semi-quantitative level, by a very simple model of resonant
transmission through a single localized state. We present our
conclusions in Sec.~\ref{sec:concl}. In Appendix \ref{appA}, we
discuss the validity of the assumption on which is based the Brownian
motion model, when it is used to describe the energy dependence of
$S$.  In this case, this amounts to study the underlying Wigner-Smith
time delay matrix. In Appendix \ref{appB} we consider the simple case
of one-channel scattering, where some exact calculations can be
carried out.

%------------------------------------------------------------------------
\subsection {Scattering matrix of a disordered region}

We consider an infinite stripe composed of two semi-infinite,
perfectly conducting regions of width $L_{y}$ (which we define as the
leads) connected by a disordered region of same width and of
longitudinal length $L_{x}$ (see the inset in
Fig.\ref{preview}.a). Assuming non-interacting electrons and hard-wall
boundary conditions for the transverse part of the wave function, the
scattering states in the leads at the Fermi energy satisfy the
condition $k^2 = (n \pi/L_{y})^2 + k^{2}_{n}$, where $k$ is the Fermi
wave vector, $n \pi/L_{y}$ the quantized transverse wave vector, and
$k_{n}$ the longitudinal wave vector. For a given $k$, each real
transverse momenta labeled by the index $n$ ($n=1,\ldots,N$) define a
propagating channel in the leads. Since each channel can carry two
waves traveling in opposite directions, in regions asymptotically far
from the scattering region the wave function can be specified by a two
$2N$-component vectors, one for each lead (labeled I and II). For both
vectors, the first (last) $N$ components are the amplitudes of the
waves propagating to the right (left). In mathematical terms, this
reads
\begin{mathletters}
\label{allwfs}
\begin{equation}
\Psi_{I}(x,y) = \sum_{n=1}^{N} \frac{1}{k_{n}^{1/2}} \left[ A_{n} e^{i
k_{n} x} + B_{n} e^{-i k_{n} x} \right] \phi_{n}(y)
\label{wfa}
\end{equation}
and
\begin{equation}
\Psi_{II}(x,y) = \sum_{n=1}^{N} \frac{1}{k_{n}^{1/2}} \left[ C_{n}
e^{i k_{n}(x-L_{x})} + D_{n} e^{-i k_{n}(x-L_{x})} \right] \phi_{n}(y).
\label{wfb}
\end{equation}
\end{mathletters}
\nin The transverse wave functions are $\phi_{n}(y)=\sqrt{2/L_{y}} \
\sin{(\pi n y/L_{y})}$. The normalization is chosen in order to have a
unit incoming flux on each channel. The scattering matrix $S$ relates
the incoming flux to the outgoing flux,
\begin{equation}
\left( \begin{array}{l}
B \\
C
\end{array} \right)
= S \
\left( \begin{array}{l}
A \\
D
\end{array} \right) .
\label{eq:Sdef}
\end{equation}
With this convention, $S$ is a $2N \! \times \! 2N$ matrix of the form
\begin{equation}
S = \left( \begin{array}{lr} r & t' \\ t & r'
\end{array} \right) .
\label{eq:Smat}
\end{equation}
The reflection (transmission) matrix $r$ ($t$) is an $N \times N$
matrix whose elements $r_{ba}$ ($t_{ba}$) denote the reflected
(transmitted) amplitude in channel $b$ when there is a unit flux
incident from the left in channel $a$. The amplitudes $r'$ and $t'$
have similar meanings, except that the incident flux comes from the
right.

The transmission and reflection amplitudes from a channel $a$ on the
left to a channel $b$ on the right and left, respectively, are given
by \cite{FishLee}
\begin{mathletters}
\label{alltrs}
\begin{equation}
t_{ba}=-i\hbar(v_{a}v_{b})^{1/2}\int dy^{\prime}\int dy \
\phi_{b}^{*}(y^{\prime}) \ \phi_{a}(y) \ G_{k}(L_{x},y^{\prime};0,y)
\label{tra}
\end{equation}
and
\begin{equation}
r_{ba}=\delta_{ab}-i\hbar(v_{a}v_{b})^{1/2}\int dy^{\prime}\int dy \
\phi_{b}^{*}(y^{\prime}) \ \phi_{a}(y) \ G_{k}(0,y^{\prime};0,y) ,
\label{trb}
\end{equation}
\end{mathletters}
\nin where $v_{a}$ ($v_{b}$) is the longitudinal velocity for the
incoming (outgoing) channel $a$ ($b$). For hard-wall boundary
conditions, $v_{\alpha}=\hbar k_{\alpha}/m$, $\alpha=a,b$. We denote
by $m$ the effective mass of the electrons. For the transmission
(reflection) amplitudes $G_{k}({\bf r'};{\bf r})$ is the retarded
Green's function between points ${\bf r} = (x,y)$ on the left lead and
${\bf r'} = (x',y')$ on the right (left) lead evaluated at the Fermi
energy $E = \hbar^2 k^2/2m$. Note that, with the convention we have
taken above, for a perfect, non-disordered sample at zero magnetic
field, $S$ is not the identity matrix, but is rather written in terms
of transmission submatrices which contain pure phases: $t_{ba}
=t'^*_{ba} = \delta_{ab}\exp{(ik_{b}L_{x})}$. Since we can pass from
one convention to the other by a fixed unitary transformation, both
forms present the same statistical properties \cite{JP}.

In our numerical work we obtain the transmission and reflection
amplitudes from the Green's function of the disordered stripe by a
recursive algorithm \cite{LeeFish,BDJS} on a tight-binding lattice. We
typically use a rectangular lattice of $34\times136$ sites and go from
the metallic to the localized regime by increasing the on-site
disorder $W$. (For details of the simulation see Ref.~\cite{JP}.)

From the transmission amplitudes one can obtain the two-terminal
conductance through the Landauer formula \cite{LandauerBuet}
\begin{equation}
g = \mbox{Tr}\ (t t^{\dagger}) .
\label{eq:Land}
\end{equation}
Here we adopt units of $e^2/h$ for the conductance. (Throughout this
work we will treat spinless electrons and therefore will not include
spin-degeneracy factors.) Notice that the Landauer formula requires
that the sample is a single, complex elastic scatterer. Thus, we
are ignoring any inelastic process giving rise to a loss of phase
coherence.

%------------------------------------------------------------------------
\subsection{Eigenphases, conductance, and Wigner time in the metallic
and localized regimes}

\begin{figure}
\setlength{\unitlength}{1mm}
\begin{picture}(200,190)
\put(0,90){\epsfxsize=13cm\epsfbox{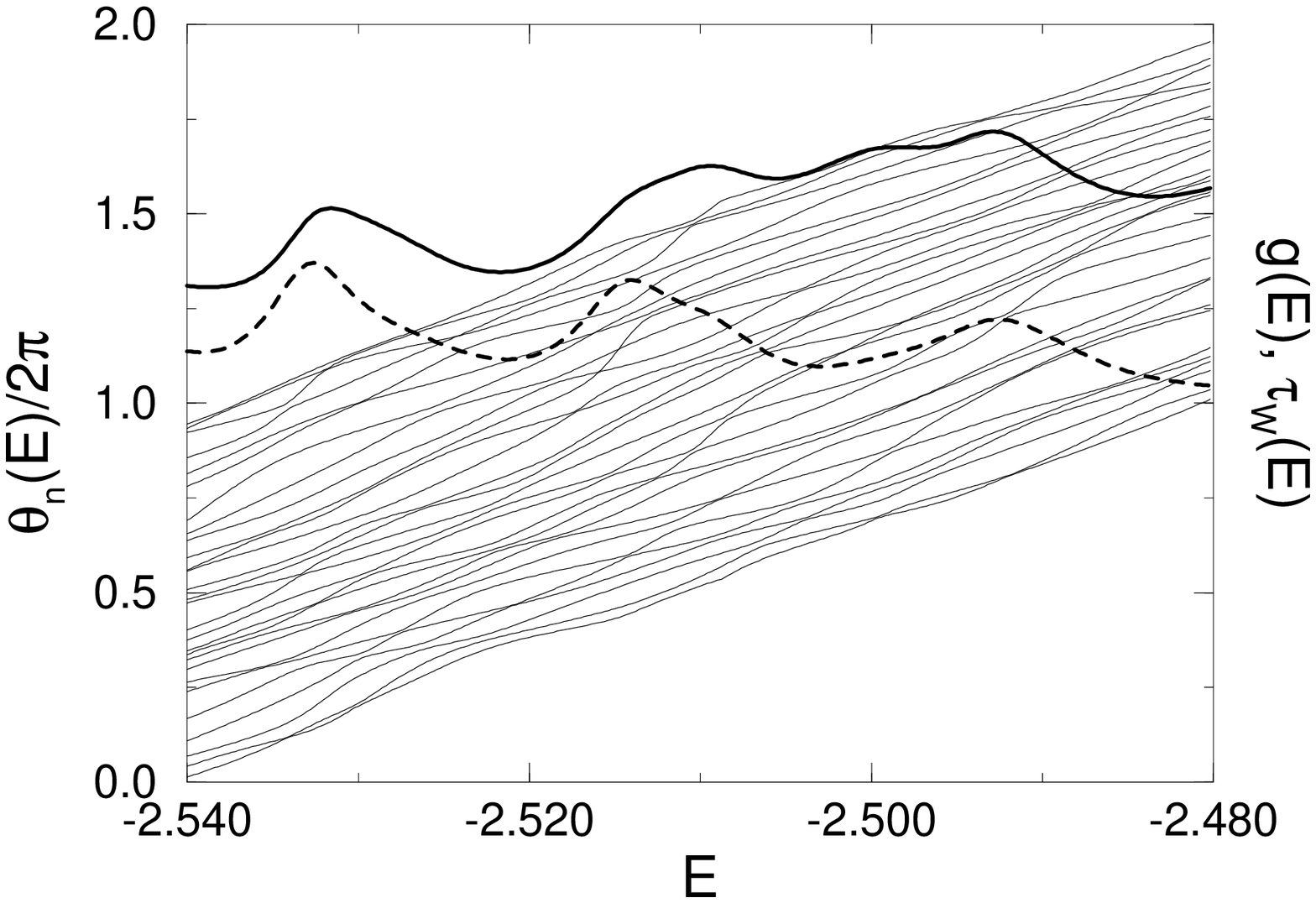}}
\put(0,0){\epsfxsize=13cm\epsfbox{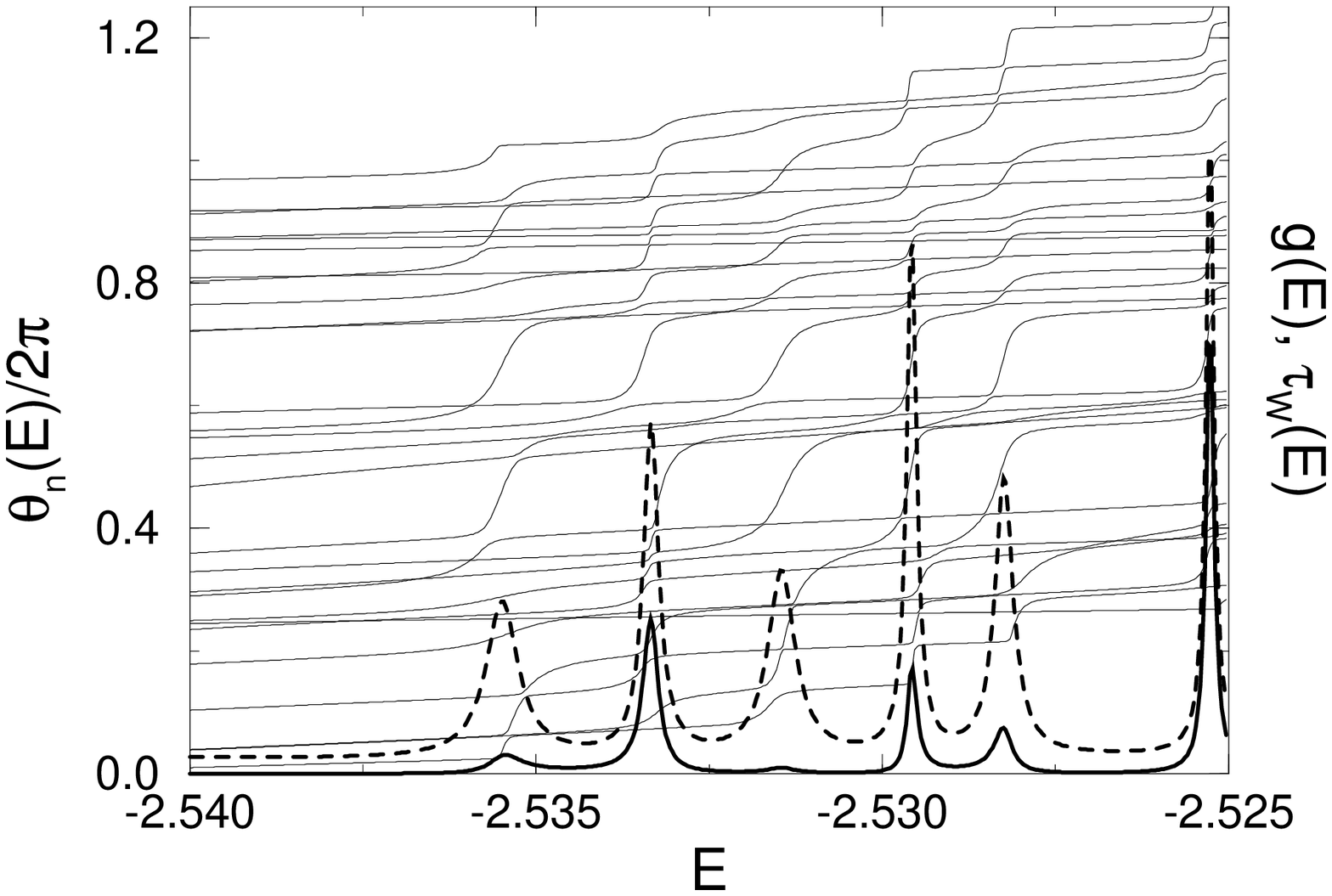}}
\put(30,170){\Large (a)}
\put(30,80){\Large (b)}
\put(62,96){\epsfxsize=4.7cm\epsfbox{fig1in.eps}}
\end{picture}
\caption{Energy dependence of the phase shifts for a quasi-1D
disordered stripe (inset) in the metallic (a) and localized (b)
regimes. The conductance and the Wigner time (thick solid and dashed
lines, respectively, both in arbitrary units) are smooth functions in
the metallic regime and exhibit a resonant structure in the localized
regime. The energy range in (b) is chosen to facilitate the visualization
of the resonances. Note the strong correlation between $g$ and $\tw$.}
\label{preview}
\end{figure}

For a given sample (i.e., impurity configuration) the diagonalization
of the scattering matrix leads to $2N$ phase shifts $\{\theta_l\}$ as
functions of the Fermi energy $E$. The typical dependence is shown in
Fig.~\ref{preview} for the metallic (a) and localized (b) cases for an
energy interval where new channels are not open ($N=14$ in the whole
interval). In the metallic case the on-site disorder in Anderson units
is $W\!=\!1$, yielding a mean-free-path $l\!=\!0.2L_{x}$, a localization
length $\xi\!=\!3 L_{x}$, and a conductance (thick solid line) which
fluctuates around a mean value $\langle g \rangle \! = \! 4.14$.

The localized regime ($W\!=\!4$, yielding $l\!=\!0.02 L_{x}$ and
$\xi\!=\!0.3 L_{x}$) exhibits a markedly different behavior,
with phase shifts showing step-like jumps where the conductance has
peaks. The peaks indicate that, for certain energies, the probability of
traversing the disordered region is much higher than the average.
They are related to the existence of localized eigenstates in the sample
and transport through the strongly-disordered region is
dominated by resonant tunneling.

To illustrate this last point, in Fig.~\ref{preview} we also show the
Wigner time for both metallic and localized regimes (thick dashed
lines). This characteristic time scale is defined as the trace of the
Wigner-Smith matrix
\begin{equation}
Q =  \frac{\hbar}{2iN} \ S^{\dagger}(E) \ \frac{dS(E)}{dE} ,
\label{WSmatrix}
\end{equation}
namely,
\begin{equation}
\tw \equiv \mbox{Tr}\ (Q) .
\end{equation}
The unitarity of $S$ trivially implies the fact that $Q$ is Hermitian, and
therefore its eigenvalues are real. In the case of time-reversal symmetry
$S^{\rm T}=S$, but $Q$ is not necessarily symmetric since in general
$S^{\dagger}$ and $dS/dE$ do not commute. Working in a base that
diagonalizes $S$ it is easy to see that $\tw$ admits a simple expression
in terms of the energy derivatives of the phase shifts,
\begin{equation}
\tw(E) = \frac{\hbar}{2N} \sum_{l=1}^{2N} \frac{d \theta_{l}(E)}{dE} .
\label{tauwapp}
\end{equation}
The Wigner time can be interpreted as the typical time interval a
scattered particle remains in the disordered region \cite{LSSS,JP}.
In the localized regime, the Wigner time exhibits the same resonant-like
structure of the conductance, although peak heights can be relatively
different. This behavior can be understood in the light of the
resonant-tunneling mechanism, since each localized state present in
the disordered region can trap the electrons for a long time \cite{KhmMuz}.
From this mechanism one can also understand qualitatively why a Wigner-time
peak can be relatively large when, at the same energy, a conductance peak is
small. This may happen, for instance, when the tunneling probability
rates for channels in lead I is much larger than for channels in lead
II. We will get back to this discussion in Sec.~\ref{sec:localized}.

A strong correlation between $g$ and $\tw$ is also obtained in the
metallic regime. The correlation in this case is also intuitive,
since transport should probe the available density of states around
the considered Fermi energy.

%%%%%%%%%%%%%%%%%%%%%%%%%%%%%%%%%%%%%%%%%%%%%%%%%%%%%%%%%%%%%%%%%%%%%%%%

\section{Correlations in the metallic regime}
\label{sec:metallic}

%------------------------------------------------------------------------
\subsection{Parametric correlations of eigenphases}
\label{sec:correl}

In order to characterize the parametric dependence on energy of the
set $\{\theta_l\}$, we define the {\it eigenphase} velocity correlator
function
\begin{equation}
C_\theta(\Delta E) \equiv \left( \frac{N}{\pi} \right)^2 \left[
\left\langle \frac{d\theta_l(E+\Delta E)}{dE} \frac{d\theta_l(E)}{dE}
\right\rangle - \left\langle \frac{d\theta_l(E)}{dE} \right\rangle^2
\right] ,
\label{eq:corphase}
\end{equation}
together with the rescaling
\begin{mathletters}
\label{eq:rescaltheta}
\begin{eqnarray}
x &\equiv & \Delta E\ \sqrt{C_\theta(0)}\\ c_\theta(x) & \equiv &
C_\theta(\Delta E)/C_\theta(0) .
\end{eqnarray}
\end{mathletters}
The average $\langle \cdots \rangle$ can be performed over different
eigenstates $n$, over the energy $E$, or over different realizations
of disorder. These definitions are analogous to the well-studied case
\cite{Aaron} in which a Hamiltonian $H$ and its eigenvalues
$\{\varepsilon_\nu\}$ depend on an external parameter $\phi$ (say, a
magnetic flux) and the {\it eigenenergy} velocity correlator is given
by
\begin{equation}
C_\varepsilon(\Delta\phi) \equiv \frac{1}{\Delta^2} \left[
\left\langle \frac{d\varepsilon_\nu(\phi+\Delta\phi)}{d\phi}
\frac{d\varepsilon_\nu(\phi)}{d\phi} \right\rangle - \left\langle
\frac{d\varepsilon_\nu(\phi)}{d\phi} \right\rangle^2 \right] ,
\label{eq:corenergy}
\end{equation}
where $\Delta$ denotes the mean level spacing, or inverse density of
states around the $\nu$-th eigenvalue. For chaotic systems, the
universal form of this correlator was checked numerically from exact
diagonalizations of suitable Hamiltonians \cite{Aaron,Ben} and after
the rescaling
\begin{mathletters}
\label{eq:rescalvarepsilon}
\begin{eqnarray}
x & \equiv & \Delta\phi\ \sqrt{C_\varepsilon(0)}\\ c_\varepsilon(x) &
\equiv & C_\varepsilon(\Delta \phi)/C_\varepsilon(0).
\end{eqnarray}
\end{mathletters}
A complete analytical expression for $c_\varepsilon(x)$ is not
available. However, the small and large-$x$ asymptotic limits are
known exactly from diagrammatic and non-perturbative calculations
\cite{Aaron,Ben} and match accurately the numerical results. In
particular, one finds that
\begin{equation}
c_\varepsilon(x) \longrightarrow -
\frac{2}{\beta(\pi x)^2} \hspace{1cm} {\rm for} \hspace{1cm}
x\rightarrow\infty ,
\label{eq:cepsasymp}
\end{equation}
where $\beta=1(2)$ for spinless systems with preserved (broken)
time-reversal symmetry.

\begin{figure}
\setlength{\unitlength}{1mm}
\begin{picture}(200,100)
\put(0,0){\epsfxsize=13cm\epsfbox{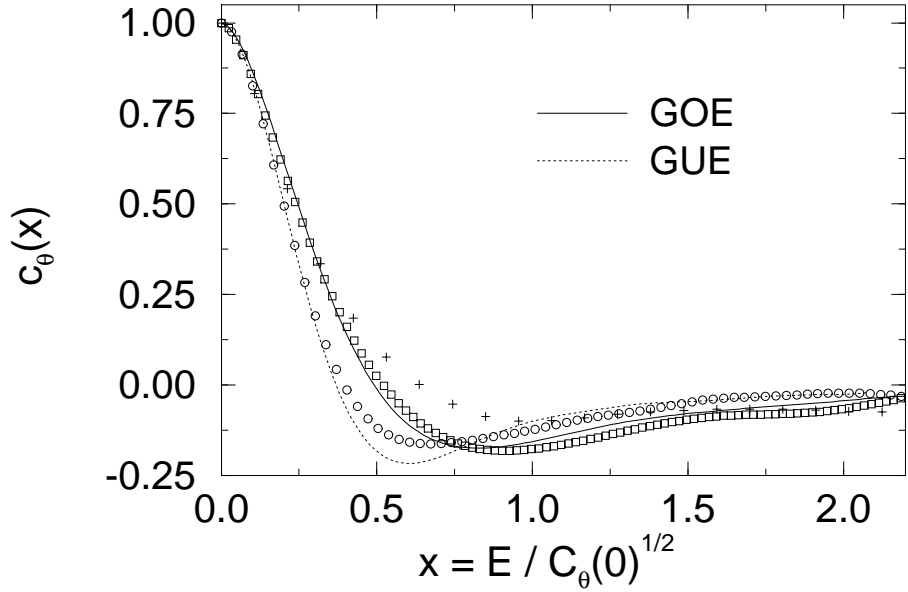}}
\end{picture}
\caption{Eigenphase velocity correlator for the metallic regime
according to the definitions of Eqs.~(\protect\ref{eq:corphase}) and
(\protect\ref{eq:rescaltheta}). The weakly-disordered metallic case
($W$=1 in Anderson units) with (circles) and without magnetic field
(squares) shows a good agreement with the universal curves
characteristic of the GUE and GOE, respectively. Increasing the disorder
($W$=2, no magnetic field, pluses) reduces the range of agreement with
the universal curve.}
\label{universal}
\end{figure}

In Fig.~\ref{universal} we present the eigenphase velocity correlation
[Eq.~(\ref{eq:corphase})] resulting from our numerical simulations and
the rescaling (\ref{eq:rescaltheta}). For weakly-disordered, metallic
samples where the statistics of the eigenphases at a fixed energy is
well described by the Dyson circular ensembles \cite{JP}, we obtain a
good agreement with the universal parametric correlation found in
Hamiltonian systems \cite{Aaron,Ben}. Applying a magnetic field
perpendicularly to the stripe, one breaks the time-reversal symmetry,
causing the parametric correlation behavior to go from GOE-like
(squares) to GUE-like (circles) \cite{curves}. Increasing the disorder
(but still remaining in the metallic regime) reduces the range of
agreement with the universal curve. Further increase of the disorder
drives the system into the localized regime and away from the
universal behavior, as we will discuss in the Section III.

The system-independent form of parametric correlations for energy
eigenvalues of random Hamiltonians has been studied with a non-linear
$\sigma$ model \cite{BenPRLNPB}. This treatment has been recently
extended \cite{Oded} to show that universality is a property of all
systems whose underlying classical dynamics is chaotic. For a
disordered sample, one finds that universality holds when $g \! \gg \! 1$
and the regime is metallic. The same approach has also been used to study
the statistical fluctuations of the $S$ matrix \cite{Macedo,FS} and
the conductance \cite{Weigr} of chaotic systems under the influence of
a {\it generic} external parameter. However, there has been no attempt
to prove analytically the results shown in Fig.~\ref{universal},
namely, that the {\it energy} correlator of eigenphase velocities
falls into the analogous curve obtained from energy eigenvalues when
the number of channels is very large.

Parallel to the field-theoretical approach, the universality of
parametric correlators of eigenvalues has also been justified from the
hypothesis of a Brownian motion of the eigenenergies \cite{Been3},
with the external parameter playing the role of a fictitious
time. The Brownian-motion model (BMM) for Hermitian and unitary matrices
was introduced by Dyson \cite{Dy2,Mehtabook} and used by Pandey
and collaborators \cite{Pandey} in the case of circular ensembles
in order to determine the statistics of the eigenphases at the
crossover between different symmetry classes. More recently,
the BMM has been applied to scattering and transmission
matrices describing coherent transport through chaotic and
disordered systems \cite{Rau,FP}. It has been recognized in these
works that the BMM approach for scattering matrices only obtains
correct results for a restricted (energy or magnetic field) range,
or for sufficiently large number of channels. We will illustrate
this point in the next subsection, where we consider specifically
the energy evolution of the phase shifts.

%------------------------------------------------------------------------
\subsection{Brownian-motion model of $S$ and energy-dependent
parametric correlators}

The aim of this subsection is to develop a simple description of the
parametric statistics of phase shifts in the metallic regime
numerically investigated ($N \! \gg \! 1$). For this purpose, we apply
a Brownian-motion model to the energy evolution of scattering
matrices. As stated at the end of the previous subsection, we do not
expect to obtain a complete quantitative agreement between the BBM
predictions and the numerical data - this certainly would require a
more sophisticated treatment \cite{VerbWeidZirn}. Instead, we only give
a justification for the agreement observed between the asymptotics of
the energy-dependent eigenphase correlation functions and analogous
curves characteristic of Hamiltonian eigenvalues.

A unitary matrix $S$ can always be decomposed as the product
\begin{equation}
S = Y^{\prime} \ Y
\label{eq:Sdecomp}
\end{equation}
of two unitary matrices. In the absence of time-reversal symmetry
($\beta=2$), $S$ is just unitary and $Y$ and $Y^{\prime}$ are independent.
For time-reversal symmetric systems ($\beta=1$), $S=S^{\rm T}$ and we
have $Y^{\prime}=Y^{\rm T}$, where the matrix $Y$ is not unique, but
specified up to an orthogonal transformation. Any permissible small
change in $S$ is then given by
\begin{equation}
\delta S = Y^{\prime} \ (i\delta \HT) \ Y ,
\label{eq:dS}
\end{equation}
where $\delta \HT$ is a Hermitian matrix (real symmetric if $\beta=1$).
This relationship allows us to define an invariant measure in the manifold
of unitary matrices from the real independent components of $\delta \HT$
\cite{Dy,Mehtabook}. The isotropic Brownian motion of $S$ occurs when $S$
changes to $S+\delta S$ as some parameter (say, a fictitious time) varies
from $t$ to $t+\delta t$. To construct the model, the components of
$\delta \HT$ are assumed to be independent random variables behaving
according to
\begin{mathletters}
\label{alleq:momQs}
\begin{equation}
\langle \delta \HT_{\mu} \rangle = 0
\label{eq:momQ1}
\end{equation}
and
\begin{equation}
f \langle \delta \HT_{\mu} \delta \HT_{\nu} \rangle = g_{\mu} \
\frac{\delta t}{\beta} \ \delta_{\mu \nu} ,
\label{eq:momQ2}
\end{equation}
\end{mathletters}
with $g_{\mu}=1+\delta_{ij}$ and $\mu = 1,\ldots, 2N + N(2N-1)\beta$.
As it will be discussed in Appendix \ref{appA}, when the evolution of
$S$ results from a Fermi energy variation $\delta E$, the Hermitian
matrix $ \delta \HT$ can be given in terms of the Wigner-Smith time
delay matrix $Q$. Thus, the validity of this Brownian motion model
relies on certain assumptions concerning the statistical properties
of the various interaction times. A random matrix approach for $Q$,
assuming a maximum entropy distribution for a given mean density of
eigenvalues, is proposed and some of its consequences are numerically
checked.  Eqs.~(\ref{alleq:momQs}) are then based on certain simplifications
which we critically discuss in Appendix \ref{appA}, at the light of
the statistical properties of the Wigner-Smith matrix $Q$.

The effect of the Brownian motion on the eigenphases of $S$ may be found by
second-order perturbation theory,
\begin{equation}
\delta \theta_n = \delta \HT_{nn}  +
\frac{1}{2} \sum_{m\ne n} (\delta \HT_{mn})^2 \ \cot
\left( \frac{\theta_n - \theta_m}{2} \right) .
\label{eq:thetaDy}
\end{equation}
Equations~(\ref{alleq:momQs}) and (\ref{eq:thetaDy}) then imply that
the eigenphases follow the relations
\begin{mathletters}
\begin{equation}
\langle \delta \theta_n \rangle = \frac{1}{f} \ F(\theta_n) \ \delta
t
\end{equation}
and
\begin{equation}
\langle \delta \theta_n \delta \theta_m \rangle = \delta_{nm}
\ \frac{2}{f\beta} \ \delta t ,
\label{eq:diffs}
\end{equation}
\end{mathletters}
with $F(\theta_n) \equiv - \partial W/\partial \theta_n$ and $W =
- \sum_{n<m} \ln | 2\sin[(\theta_n-\theta_m)/2] |$. Notice that the
coefficients $f$ and $\beta$ were introduced to suggest friction and
inverse temperature, respectively. The eigenphases behave like a classical
gas of massless particles on the unit circle, executing a Brownian motion
under the influence of a Coulomb force $F$. The joint probability
distribution of the eigenphases $P(\{\theta_n\};t)$ follows a Fokker-Planck
equation \cite{Dy2,Mehtabook}
\begin{equation}
\frac{\partial P}{\partial t} = \frac{1}{f} \ \sum_{n=1}^{2N}
\frac{\partial}{\partial \theta_n} \left( \frac{1}{\beta}
\frac{\partial P}{\partial \theta_n} + P \frac{\partial W}{\partial
\theta_n} \right) .
\label{FokkerPlanck}
\end{equation}
For $t\rightarrow\infty$ we reach an equilibrium situation
where the joint distribution becomes that of the circular ensemble
\cite{Mehtabook}, $P_{eq}(\{\theta_n\}) = C_{2N} \exp[-\beta
W(\{\theta_n\})]$.

We now turn to the solution of the Fokker-Planck equation for the
joint probability distribution of the eigenphases. An exact
way to solve Eq.~(\ref{FokkerPlanck}) is to map it into a
quantum-mechanical Hamiltonian problem of particles in a ring
interacting by a two-body potential proportional to $1/r^2_c$, where
$r_c^2$ is the length of the chord connecting the pair. This model was
proposed and studied extensively by Sutherland \cite{Sutherland} and
yields an exact solution for $P(\{\theta_n\};t)$ \cite{BeenRejaei}.
Alternatively, one may use a hydrodynamical approximation, originally
proposed by Dyson \cite{Dy3}, which leads to a non-linear diffusion
equation for the average density of eigenphases. Here we will adopt this
second approach, since it is simpler than the first one and sufficient
for our purposes. Thus, introducing the $t$-dependent density of
eigenphases,
\begin{equation}
\rho(\theta;t) \equiv \int_0^{2\pi} d\theta_1 \cdots d\theta_{2N}
P(\{\theta_l\};t) \sum_{n=1}^{2N} \delta (\theta - \theta_n),
\end{equation}
and starting from Eq.~(\ref{FokkerPlanck}), one can derive that
\begin{equation}
f \frac{\partial\rho(\theta;t)}{\partial t} = -
\frac{\partial}{\partial t} \left[ \rho(\theta;t)
\frac{\partial}{\partial \theta} \int_0^{2\pi} d\theta^\prime
\rho(\theta^\prime;t) \ln \left |2 \sin \left(
\frac{\theta-\theta^\prime}{2} \right) \right| \right].
\end{equation}
(This equation is approximate in the sense that it does not take into
account accurately short-wavelength oscillations in the density.)
Since fluctuations in the density are smaller than the density itself
by a factor of the order $O(1/N)$, one can linearize this diffusive
equation by considering small deviations around the homogeneous
solution, $\rho(\theta;t) = \rho_{eq} + \delta \rho(\theta;t)$, where
$\rho_{eq}= N/\pi$. One then obtains
\begin{equation}
\frac{\partial \delta \rho (\theta;t)} {\partial t} =
\frac{\partial}{\partial \theta} \int_0^{2\pi} d\theta^\prime
D(\theta-\theta^\prime) \frac{\partial \delta \rho (\theta^\prime;t)
}{\partial \theta^\prime},
\end{equation}
where the kernel is given by $D(\theta) = - \rho_{eq} f^{-1} \ln
|2 \sin [(\theta-\theta^\prime)/2]|$. The periodicity in the
eigenphase space calls for a solution of the diffusion equation in
Fourier series. Writing $\delta \rho(\theta;t) =
\sum_{k=-\infty}^{+\infty} \delta \rho_k(t) e^{ik\theta}$, one finds
that
\begin{equation}
\delta\rho_k(t) = \delta\rho(0) \exp
\left(-\frac{\pi|k|\rho_{eq}t}{f} \right).
\label{eq:Fourdens}
\end{equation}
The above result can be readily used in the evaluation of parametric
correlators, which will be the subject of the rest of this
subsection. We therefore introduce the two-point density correlator
\begin{equation}
R(\theta,\theta^\prime;t,t^\prime) \equiv \langle \delta
\rho(\theta;t) \delta \rho(\theta^\prime;t^\prime) \rangle_{eq} ,
\end{equation}
where $\langle \cdots \rangle_{eq}$ means that the average is weighted
by the joint distribution at the equilibrium,
$P_{eq}(\{\theta_l\})$. Since the average density is constant in both
time and angle, there is translation invariance in $t$ and $\theta$ at
the equilibrium. Consequently, $R$ depends only on the differences
$\theta - \theta^\prime$ and  $t - t^\prime)$. From
Eq.~(\ref{eq:Fourdens}), we find that the dependence on time and angle
can be decoupled in $k-$space,
\begin{equation}
R_k(t) = T_k \exp \left( \frac{-\pi|k| \rho_{eq} t}{f} \right),
\label{eq:coeffR}
\end{equation}
where $T_k$ is the Fourier coefficient of the ``static'' correlator
$T(\theta) \equiv R(\theta;t=0)$, which, for circular ensembles, is
known exactly \cite{Mehtabook}. In the limit of $N\gg1$, one has that
$T_k \simeq |k|/(2\pi^2\beta)$ for $|k|\ll N$, whereas for $|k|\gg N$
it saturates at the value $T_k \simeq N/(2\pi)^2$. If we keep $\pi
\rho_{eq} t/f \gg 1/N$, we can neglect the contribution of terms
with $|k|>N$ and then easily sum the Fourier expansion of
$R(\theta;t)$ to find that
\begin{equation}
R(\theta;t) \simeq \frac{1}{\pi^2\beta} \mbox{Re} \left[
\frac{z}{(1-z)^2} \right],
\label{eq:denscorrl}
\end{equation}
with $z = \exp(i \theta - \pi \rho_{eq} t /f)$. Hereafter we will
denote the energy (or time) in terms of a dimensionless parameter,
with the natural choice being $X \equiv E/E_s$, where $E_s =
\sqrt{f/(\pi\rho_{eq})}$.

The knowledge of $R(\theta;t)$ permits the evaluation of other
two-point parametric functions. Two of them are of particular
interest. The first one is the Wigner time correlator
\begin{eqnarray}
\label{eq:taucor}
C_\tau(\Delta E) & \equiv & \langle \tw(E+\Delta E)\tw(E) \rangle -
\langle \tw(E) \rangle^2 \nonumber \\
& = & \left(\frac{\hbar}{2N}\right)^2 \
\sum_{l,m=1}^{2N} \left[ \left\langle \frac{d\theta_l(E+\Delta E)}{dE}
\frac{d\theta_m(E)}{dE} \right\rangle - \left\langle
\frac{d\theta_l(E)}{dE} \right\rangle \left\langle
\frac{d\theta_m(E)}{dE} \right\rangle \right] ,
\end{eqnarray}
which has also been the subject of recent calculations
\cite{LSSS,FS}. The second parametric function is the modified level velocity
correlator \cite{Ben,Been3}
\begin{equation}
\tilde{C}(\theta-\theta^\prime;E-E^\prime) \equiv \sum_{n,m=1}^{2N}
\left\langle \frac{d\theta_n (E)}{dE} \frac{d\theta_m
(E^\prime)}{dE^\prime} \delta \biglb( \theta - \theta_n(E) \bigrb)
\delta \biglb( \theta^\prime - \theta_m(E^\prime) \bigrb)
\right\rangle_{eq}.
\end{equation}
It is straightforward to check that
\begin{equation}
C_\tau(E) = -\frac{\pi}{(2N)^2} \frac{\partial^2}{\partial E^2}
\int_0^{2\pi} d\theta\ (\theta - \pi)^2 R(\theta;E)
\label{eq:eqvCC}
\end{equation}
and
\begin{equation}
\frac{\partial^2 R}{\partial E^2} = \frac{\partial^2
\tilde{C}}{\partial \theta^2}.
\label{eq:relSC}
\end{equation}
In Eq.~(\ref{eq:eqvCC}) one uses that $R(\theta;E)$ is an even
periodic function in $\theta$ and $R_k(E)=0$ for $k=0$.] Moreover, the
two correlators are connected by the relation
\begin{equation}
C_\tau(E) = \frac{2\pi}{(2N)^2} \int_0^{2\pi} d\theta\
\tilde{C}(\theta;E),
\end{equation}
which also implies that $\tilde{C}(0;E) = 2(N/\pi)^2 C_\tau (E)$.

Two remarks are pertinent here. First, $C_\tau(E)$ and
$\tilde{C}(0;E)$ have asymptotic forms similar to $C_\theta(E)$ in the
limit of large $E$. This is because distinct eigenphases become
uncorrelated very quickly for large enough energy differences, causing
the main contribution to both correlators to come from the diagonal
terms. The second remark is that, independently of the hydrodynamic
approximation adopted above, interlevel correlations are completely
ignored in the BBM [see Eq.~(\ref{eq:diffs})]. This leads to a $C_\tau(E)$
identical, up to a proportionality factor, to the one-level velocity
correlator $C_\theta(E)$ {\it for all values of $E$}. Such a coincidence
between $C_\tau(E)$ and $C_\theta(E)$ is never observed in practice
because interlevel correlations are indeed important, particularly
among neighboring levels and at small values of $E$.

From Eqs.~(\ref{eq:coeffR}), (\ref{eq:eqvCC}), and (\ref{eq:relSC}) one
gets
\begin{equation}
C_\tau(E) = -\frac{1}{2\beta N^2 E_s^2} \frac{ (e^{-X^2} - 1 + 2X^2)}
{\sinh^2 (X^2/2)}
\label{eq:ctaubmm}
\end{equation}
and
\begin{equation}
\tilde{C}(\theta;E) = \left(\frac{N}{\pi} \right)^2 C_\tau(E) -
\frac{2}{\pi^2\beta E_s^2} \mbox{Re} \left[ \frac{z^2 -
z(1-2X^2)}{(1-z)^2} \right] .
\label{eq:ctildebmm}
\end{equation}
We find the following asymptotic limits: ({\it i}) For $X\!\ll\!1$,
$C_\tau(E) \simeq -2/(\beta N^2 E^2)$ and $\tilde{C}(0;E) \simeq
-4/(\pi^2 \beta E^2)$; ({\it ii}) for $X\!\gg\!1$, $C_\tau(E) \simeq
-4X^2 e^{-X^2}/ (\beta N^2 E_s^2)$ and $\tilde{C}(0;E) \simeq
-8X^2 e^{-X^2}/ (\pi^2 \beta E_s^2)$. An exponential decay at large
distances does not occur for the analogous correlators involving
eigenvalue velocities \cite{Been3}, which actually retains the form
({\it i}) for arbitrarily large values of the external parameter.
As pointed out before \cite{FP}, technically, the difference appears
because a Fourier series is used to treat phase shifts oscillations
(since they are bounded by the finite interval $[0;2\pi]$) instead
of Fourier transforms, as in the case of energy eigenvalues.
Nevertheless, if $N \! \gg \! 1$, one naturally expects that there
should exist a range of values of $E$ in which $\tilde{C}(\theta;E)$
has a form similar to its counterpart for eigenenergies. In order to
understand this point, let us find the relation between $E_s$ and
another, more commonly used scale, $C_\theta (0)^{-1/2}$ \cite{Aaron,Ben}
(see Sec.~\ref{sec:correl}). From Eq.~(\ref{eq:diffs}) we have that
\begin{equation}
C_\theta (0)^{-1/2} = \frac{1}{\rho_{eq}} \sqrt{
\frac{f\beta}{2}}.
\label{eq:c0beta}
\end{equation}
which means that $E_s = C_\theta(0)^{-1/2} \sqrt{2N
/(\pi^2\beta)}$. Therefore, if $N$ is large enough, it is possible to
have $E\gg E_s$ and $E\ll C_\theta(0)^{-1/2}$ simultaneously. Consequently,
there is an intermediate range of values of $X$ where $\tilde{C}(0;X)$
falls exactly into the same power-law, universal curve predicted for the
case of energy eigenvalues \cite{Ben,Been3}. Deviations in the form of
an exponential decay occur only for values of $X \gtrsim 1$, which may
be large in units of $C_\theta(0)^{-1/2}/E_s$. Notice that this is still
compatible with the restriction $E/E_s\gg 1/N$ used to derive
Eq.~(\ref{eq:denscorrl}).

We stress again that Eqs.~(\ref{eq:ctaubmm}) and (\ref{eq:ctildebmm})
are approximations. A simple inspection is sufficient to prove this point:
Because the correlators diverge at $E=0$ as $1/E^2$, they cannot satisfy the
sum rules $\int_0^\infty dE\ \tilde{C}(\theta;E) = 0$ (for any $\theta$)
and $\int_0^\infty dE\ C_\tau(E) = 0$. This limitation cannot be overcome
within the hydrodynamical approximation we have considered here. In fact,
Eqs.~(\ref{eq:ctaubmm}) and (\ref{eq:ctildebmm}) break down for $X \lesssim
C_\theta(0)^{-1/2}$. To go beyond this limitation, one has to treat
Eq.~(\ref{FokkerPlanck}) in a non-perturbative way.

There is an additional, appealing relation connecting $E_s$ to another
scale inherent to the scattering region. Recall that $E_s =
(\hbar/N)/ \sqrt{\beta (\langle \tw^2 \rangle - \langle \tw \rangle^2)}$.
The variance of the Wigner time was evaluated in Refs.~\onlinecite{LSSS,FS}
using a microscopic formulation to relate $S$ to the Hamiltonian of the
scattering region, which was modeled as a member of a Gaussian ensemble
\cite{VerbWeidZirn}. (We remind the reader that Gaussian ensembles are
supposed to model the statistical properties of ballistic chaotic
cavities as well as disordered electronic systems in the diffusive regime.)
It was shown that
\begin{equation}
\langle \tw^2 \rangle = \langle \tw \rangle^2 \left[ 1 + \frac{4}{\beta
(2N)^2} \right]
\label{eq:Fyod}
\end{equation}
when the scattering region is maximally connected to the external propagating
channel (i.e., open leads) and $N\gg 1$. Based on this result, we infer
\begin{equation}
E_s = \frac{N\Delta}{\pi}.
\label{eq:relat}
\end{equation}
This relation says that in the metallic regime of a quasi-1D wire, only
one fundamental energy scale, the mean level spacing, is required to
characterize the decay of energy-parametric correlators. Other
system-dependent scales, like the Thouless energy $E_c = \hbar v_F l/L^2$
are irrelevant. One should recall that, for energy levels, the analogous
quantity to $C_\theta(0)^{-1/2}$ is the root-mean square velocity
$C_\varepsilon(0)^{-1/2}$ [see Eq.~(\ref{eq:rescalvarepsilon})], which
is a direct measure of the average dimensionless conductance $\langle
g\rangle \sim E_c/\Delta$ {\it when the system is in the metallic regime}
\cite{Ben,Akkermans92}. There is no direct relation between
$C_\theta(0)^{-1/2}$ and $C_\varepsilon(0)^{-1/2}$ and one cannot recover
any quantitative information about the conductance of the system by
calculating $C_\theta(E)$ alone in the metallic regime.

In Fig.~\ref{metcorr} we present the Wigner time correlator
(\ref{eq:taucor}) with the energy rescaled according to
Eq.~(\ref{eq:rescaltheta}). Notice that the shape of $C_{\tau}$
is very similar for different symmetry classes ($\beta=1$ and $2$).
This property also emerges in the context of the stochastic approach to
scattering \cite{VerbWeidZirn} when we take the large-$N$ asymptotics of
$C_{\tau}$ \cite{LSSS,FS}: The form of the correlator becomes
\begin{equation}
\label{eq:Semclas}
C_\tau(E) = C_\tau(0) \frac{1 - (E/\Gamma)^2}{[1 + (E/\Gamma)^2]^2},
\end{equation}
with $\Gamma=N\Delta/\pi$. This curve matches reasonably well
our data when we let $\Gamma$ be a {\it free} fitting parameter.

From a semiclassical point-of-view, the existence of time-reversal 
symmetry {\it in a chaotic system} does not affect the shape of any energy-dependent, two-point correlator of elements of $S$ \cite{BluSmi,BJS} 
or $\tw$ \cite{eckhardt}. This is because, after energy averaging, these 
correlators are solely determined by the exponential decay with time of 
the classical probability to escape from the scattering region, in which 
case $\Gamma^{-1}$ is interpreted as the escape rate. The presence or not 
of time-reversal symmetry affects only numerical prefactors in the 
correlators. The fact that large $N$ corresponds to a semiclassical 
limit becomes clear when we notice that taking $\hbar \to 0$ for a fixed 
lead geometry effectively increases the number of propagating 
channels \cite{caio91}.

%------------------------------------------------------------------------
\subsection{Correlation between the Wigner time and the conductance}

In the previous subsection we discussed the Brownian motion model for $S$
and the importance of the probability distribution of the time-delay matrix
(further developed in Appendix \ref{appA}). We now focus on the statistical
properties of the trace of $Q$, the Wigner time $\tw$ in the metallic regime.
As evident from Fig.\ref{preview}, there are strong correlations between
the Wigner time and the conductance, that we discuss below. We start by
writing the scattering matrix in its polar decomposition \cite{St}
\begin{equation}
S = \tilde U \ \Gamma \ U .
\label{eq:polar}
\end{equation}
The $2N\!\times\!2N$ unitary matrices $U$ and $\tilde U$ are built out of
unitary $N\!\times\!N$ blocks, namely,
\begin{equation}
U = \left( \begin{array}{cc} u^{(1)} & \ 0 \\ 0 & \ u^{(2)}
\end{array} \right)
\quad \mbox{and} \quad \tilde U = \left(
\begin{array}{cc} u^{(3)} & \ 0 \\ 0 & \ u^{(4)}
\end{array} \right) .
\label{eq:matU}
\end{equation}
For systems with broken time-reversal symmetry, the matrices $u^{(l)}$
are independent of each other. If time-reversal symmetry is preserved,
then $S$ is symmetric and one has $u^{(3)} = u^{(1)T}$ and $u^{(4)} =
u^{(2)T}$. The $2N\!\times\!2N$ matrix $\Gamma$ has the block structure

\begin{equation}
\Gamma = \left( \begin{array}{cc} -{\cal R} & \quad {\cal T} \\ {\cal
T} & \quad {\cal R}
\end{array} \right) ,
\label{eq:Spol}
\end{equation}
where ${\cal R}$ and ${\cal T}$ are real diagonal $N\!\times\!N$ whose
non-zero elements can be expressed as
\begin{equation}
\label{allrtcals}
{\cal R}_{a} = \left(\frac{\lambda_{a}}{1+\lambda_{a}}\right)^{1/2}
\hspace{1cm} \mbox{and} \hspace{1cm} {\cal T}_{a} =
\left(\frac{1}{1+\lambda_{a}}\right)^{1/2}
\end{equation}
in terms of the radial parameters $\lambda_{a}$. The convenience of
this representation is that it allows the two-probe conductance of
Eq.~(\ref{eq:Land}) to be expressed simply as
\begin{equation}
g = \sum_{a=1}^{N} \ \frac{1}{1+\lambda_{a}} ,
\label{eq:geigen}
\end{equation}
that is, independent of the unitary matrices $U$ and $\tilde U$.

In the absence of a magnetic field there is time-reversal symmetry
($S = S^{\rm T}$, \ $\tilde U = U^{\rm T}$), and we have
\begin{equation}
S^{\dagger} \ \frac{dS}{dE} = U^{\dagger} \Gamma
\left[ \frac{d \Gamma}{dE} U + \Gamma \frac{d U}{dE}
-\frac{d U^{*}}{dE} U^{\rm T} \Gamma U \right] .
\label{inter}
\end{equation}
Since $\Gamma^2=I$, taking the trace of the above equation simplifies
it, yielding
\begin{equation}
\tw(E) = - i \ \frac{\hbar}{2N} \ \mbox{Tr} \left( U^{\dagger} \frac{d
U}{dE} - \frac{d U^{*}}{dE} U^{\rm T} \right) .
\label{tauwdou}
\end{equation}
Moreover, because $U$ is unitary, its infinitesimal variations are
given by $dU = \delta U \ U$, where $\delta U$ is
antihermitian. Therefore,
\begin{equation}
\tw(E) = - i \ \frac{\hbar}{2N} \ \mbox{Tr} \left( \frac{\delta U}{dE}
- \frac{\delta U^{*}}{dE} \right) .
\label{tauwdou2}
\end{equation}
Writing for the block components of $U$
\begin{equation}
du^{(l)} = \delta u^{(l)} \ u^{(l)} \quad \mbox{and} \quad \delta
u^{(l)} = da^{(l)}+i \ ds^{(l)} ,
\label{eq:dvl}
\end{equation}
$l=1,2$, where $d a^{(l)}$ ($d s^{(l)}$) are real antisymmetric
(symmetric) $N \! \times \! N$ matrices, we have
\begin{equation}
\tw(E) =  -  \frac{\hbar}{N} \sum_{l=1}^{2} \ \sum_{a=1}^{N}
\frac{d s^{(l)}_{aa}}{dE} .
\label{tauwdou3}
\end{equation}
Notice that here $\langle \tw \rangle = 0$, since the polar
decomposition (\ref{eq:polar}) yields the identity for $S$ in the
absence of disorder. However, with the convention of
Eq.~(\ref{allwfs}) which we took for defining our scattering matrix in
the numerical simulations, we do not have $S=I$ in the absence of
disorder. As a consequence, the mean average value of the Wigner time
obtained numerically is given by the density of states of the
disordered region \cite{JP,DoSm}. However, as expressed in
Sec.~\ref{sec:intro}, when we go from one convention to another
by multiplying $S$ by a fixed unitary matrix, we do not change the
statistical properties of the eigenphases. In the same way, the
constant shift in $\tw$ given by the density of states does not change
its statistical properties.

\begin{figure}
\setlength{\unitlength}{1mm}
\begin{picture}(200,100)
\put(0,0){\epsfxsize=13cm\epsfbox{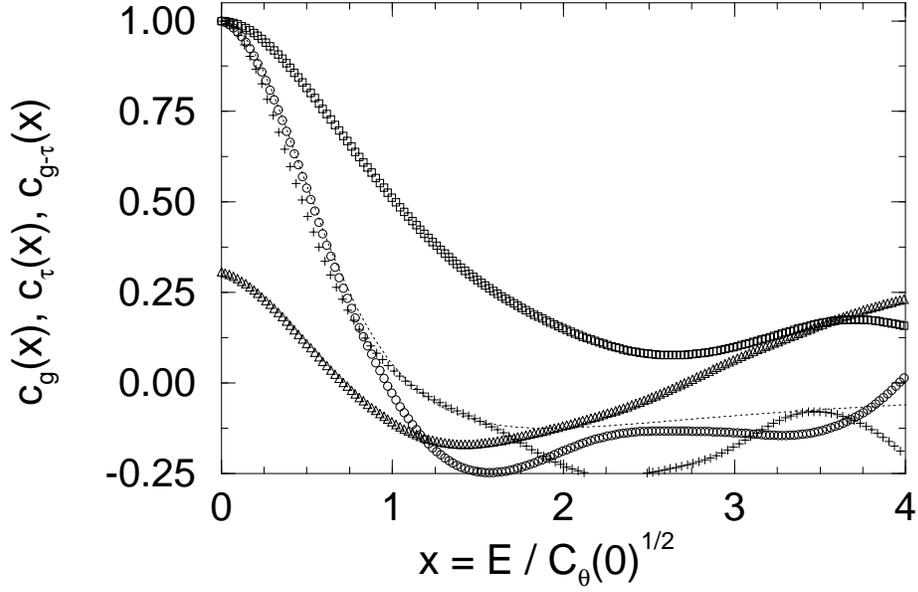}}
\end{picture}
\caption{Autocorrelations of the conductance $g$ (square) and the Wigner
time $\tw$ (circle) and the cross-correlation between $g$ and $\tw$
(triangle) for the metallic case without magnetic field. The
autocorrelator of Wigner times in the presence of a magnetic field is
also plotted (pluses). The dotted curve is a plot of
Eq.~(\protect\ref{eq:Semclas}).}
\label{metcorr}
\end{figure}

One interesting feature of Eqs.~(\ref{tauwdou2}) and (\ref{tauwdou3})
is that $\tw$ only depends on the infinitesimal variations of the
unitary matrix $U$ and not on the radial parameters $\lambda$ of the
polar decomposition. This may seem somehow surprising, given the
obvious correlations between the Wigner time and the conductance [as
shown in Figs.~\ref{preview}.a and \ref{metcorr}] and the fact that the
latter depends exclusively on the radial parameters through
Eq.~(\ref{eq:geigen}). However, when the electron Fermi energy is
varied for a given impurity configuration, the corresponding changes
in $\lambda$ and $\delta u^{(l)}$ are necessarily correlated since $S$ is
constrained by symmetry to move along a particular direction in the
manifold of allowed polar parameters. In other words, since the Wigner
time is given by the energy derivative of $S$, it must comply with the
symmetry requirements in the same way as the infinitesimal variations
that define the invariant measure.

In Fig.~\ref{metcorr} we present the $g-\tau$ correlator
\begin{equation}
C_{g\tau}(\Delta E) = \frac{\langle \delta g (E + \Delta E) \ \delta
\tw (E) \rangle} {\langle (\delta g)^2 \rangle^{1/2} \langle (\delta
\tw )^2 \rangle^{1/2}} ,
\label{eq:corrgt}
\end{equation}
together with the $g-g$ and $\tau-\tau$ correlators. (Here we define
$\delta \tw \equiv \langle \tw^2 \rangle - \langle \tw
\rangle^2$. Analogously for $\delta g$.) The correlator of Wigner
times has been recently calculated by several authors \cite{LSSS,FS}
using supersymmetry methods, while the $g-g$ correlator is only known
analytically in the metallic perturbative regime \cite{lee1} when $E
\gg \Delta$. The calculation of the $g-\tau$ correlator will be
very interesting, given the numerical evidence provided.

%%%%%%%%%%%%%%%%%%%%%%%%%%%%%%%%%%%%%%%%%%%%%%%%%%%%%%%%%%%%%%%%%%%%%%%%%
\section{Correlations in the localized regime}
\label{sec:localized}

%------------------------------------------------------------------------
\subsection{Resonant transmission model for localized transport}

As evident from Fig.~\ref{preview}, transport through a localized
stripe presents important differences for the eigenphases, conductance,
and Wigner time as compared to the metallic case of Section
\ref{sec:metallic}. The peaks of $g(E)$ and $\tw(E)$ and the jumps of
$\theta_l(E)$ show that now we are in a resonant regime, where the
transmission occurs through tunneling into localized eigenstates in
the bulk of the disordered region. The dependence of transport properties on
resonant states can be established within the R-matrix formalism
\cite{LaTo}, which allows the scattering matrix to be expressed as
\cite{Mahaux69} (see Appendix \ref{appB})
\begin{equation}
S_{nm}(E) \simeq \delta_{nm} - 2\pi i\sum_{\nu} \frac{W^\ast_{n\nu}
W_{m\nu}} {E - E_\nu + i \Gamma_\nu/2} .
\label{eq:Smatrixloc}
\end{equation}
The sum is over the (localized) eigenstates of the disordered region,
the matrix elements $W_{m\nu}$ describe the coupling of these states
with the different channels in the leads, and $\Gamma_\nu =
2\pi\sum_{n=1}^{2N} |W_{n\nu}|^2$ is the resonance total width for the
eigenstate $\nu$. Equation~(\ref{eq:Smatrixloc}) is valid only to
lowest order in $\Gamma_{\nu}/\Delta$, namely, when resonances do not
overlap. (Higher order corrections imply traversing the disordered
region by sequential tunneling through more than one localized state.)
Within this approximation, the energy dependence of the conductance
and the Wigner time appears in the form of Breit-Wigner functions
\begin{equation}
\label{gBW}
g(E) \simeq \sum_{\nu}
\frac{\Gamma_\nu^{(l)}\Gamma_\nu^{(r)}} {(E-E_\nu)^2+\Gamma_\nu^{2}/4}
\end{equation}
and
\begin{equation}
\label{twBW}
\tw(E) \simeq \frac{\hbar}{2N} \ \sum_{\nu}
\frac{\Gamma_\nu} {(E-E_\nu)^2+\Gamma_\nu^{2}/4} ,
\end{equation}
respectively. The left ($\Gamma_\nu^{(l)}$) and right ($\Gamma_\nu^{(r)}$)
partial widths are given by the overlap of the corresponding eigenfunction
$\psi_\nu (x,y)$ with the channel wave functions $\phi_{n}(y)$, i.e.,
\begin{equation}
\label{parwi}
\Gamma_{\nu}^{(l)} = \Delta \sum_{n=1}^{N} c_n \left| \int_{0}^{L_y}
dy\ \phi_{n}(y) \psi_{\nu}(x\!=\!0,y) \right|^2
\end{equation}
and similarly for $\Gamma_{\nu}^{(r)}$, exchanging $x=0$ by $x=L$. The
total width is $\Gamma_\nu = \Gamma_{\nu}^{(l)} + \Gamma_{\nu}^{(r)}$.
The coefficients $c_n=\hbar^2k_{n}/(2m\Delta)$ are smooth functions of
energy on the scale of $\Delta$ ($k_{n}$ is the longitudinal wave vector
defined in Sec.~\ref{sec:intro}.B). In the strongly localized case assumed
here, the typical total width $\Gamma_\nu$ is much smaller than $\Delta$
and only the eigenstate $\nu$ whose energy is the closest to $E$ contributes
significantly to the sum. Hence, from now on we will neglect any smooth
energy dependence in $\Gamma_\nu$ and omit the index $\nu$. The fitting
of $g(E)$ by a single Breit-Wigner resonance works quite well for our
numerical data when $W=4$ or larger; other numerical models of disorder
\cite{AviPicMut} also yield Breit-Wigner shapes.

The statistical properties of $\Gamma$ are connected to the
fluctuations in the eigenfunction intensity. Using some very simple
arguments one can estimate the probability distribution
$P(\Gamma)$. First we recall that the envelope of a localized state
decays as one moves away from its center $r_0=(x_0,y_0)$. The scale of
this decay is the localization length $\xi$, such that $|\psi(r)| \sim
\exp{(-|r-r_0|/\xi)}$. Consequently, we may write that
\begin{mathletters}
\label{allgs}
\begin{equation}
\Gamma^{(l)} \sim e^{-2 x_0/\xi} \ f_N^{(l)}\{\psi\}
\end{equation}
and
\begin{equation}
\Gamma^{(r)} \sim e^{-2(L-x_0)/\xi} \ f_N^{(r)}\{\psi\} .
\end{equation}
\end{mathletters}
\nin The factors $f_N$ arise from the fluctuations of the eigenstate
on the scale of $k_F^{-1}$. However, since we work with a large number
of channels $N$, $f_N$ follows approximately a Gaussian distribution and
therefore can be substituted by the its average value. Factorizing away
the energy scale given by the level spacing $\Delta$, we define
\begin{mathletters}
\label{allgds}
\begin{equation}
\gamma^{(l)} \equiv \frac{\Gamma^{(l)}}{c \Delta} \sim e^{-2x_0/\xi}
\end{equation}
and
\begin{equation}
\gamma^{(r)} \equiv \frac{\Gamma^{(r)}}{c \Delta} \sim
e^{-2(L-x_0)/\xi},
\end{equation}
\end{mathletters}
where $c$ is a numerical constant proportional to $N$. Analogously,
we define the dimensionless total width $\gamma = \gamma^{(l)} +
\gamma^{(r)}$.

Our statistical assumptions will be the following:

\begin{enumerate}
\item $x_0$ is uniformly distributed along the disordered stripe,
\begin{equation}
\label{assumpa}
P(x_0) = \frac{1}{L} ;
\end{equation}
\item $x_0$ and $\xi$ are independent random variables;
\item $z=2L/\xi$ has a normal distribution \cite{Pichard}
\begin{equation}
\label{assumpb}
P(z) = F \ \exp \left[ -\frac{(z-z_0)^2}{2\sigma^2} \right] ,
\end{equation}
\end{enumerate}
where the mean and the variance are related by $\sigma^2=2z_0$. The two
first assumptions are trivial. The third assumption originates from the
standard log-normal distribution of the dimensionless conductance
\cite{Pichard}, $g=e^{-z}$, with
\begin{equation}
\label{LN}
\mbox{var}(\ln{g}) = -2 \ \langle \ln{g} \rangle .
\end{equation}

\begin{figure}
\setlength{\unitlength}{1mm}
\begin{picture}(200,150)
\put(0,0){\epsfxsize=12cm\epsfbox{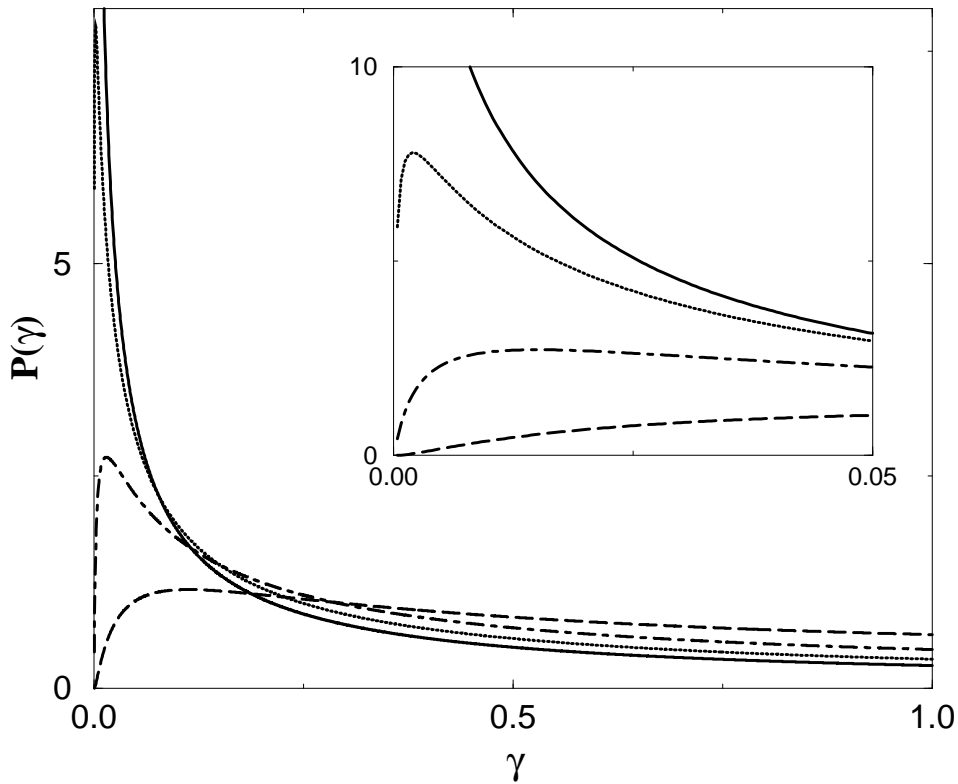}}
\end{picture}
\caption{Probability distribution for the dimensionless total width
$\gamma$ for increasing disorder (parameterized by $z_0$). Dashed line:
$z_0=4$, corresponding to the case shown in Fig.~(\protect\ref{preview}).b;
dash-dotted: $z_0=6$; dotted: $z_0=8$; solid line: $z_0=10$. Inset:
blow up of the small $\gamma$ region showing how the most probable
value of the distribution moves towards the origin with increasing
disorder.}
\label{pgammanum}
\end{figure}

The mean value $z_0$ is a measure of the disorder, and in the strongly
localized regime we have $z_0\!\gg\!1$. The normalization factor
$F = [\sqrt{\pi z_0} (1 + \Phi(\sqrt{z_0}/2))]^{-1}$ takes into account
the fact that $z$ is always positive. ($\Phi$ denotes the error
function.) Nevertheless, to leading order in $1/z_0$, we can ignore this
restriction over $z$ and recover the standard Gaussian prefactor $F
\simeq [2\sqrt{\pi z_0}]^{-1}$. The probability distribution of
$\gamma$ can be constructed as
\begin{equation}
P(\gamma) \simeq 2F \int_0^{1/2} ds \int_0^\infty dz \exp \left[
-\frac{(z-z_0)^2}{4 z_0} \right] \delta \Biglb( \gamma - 2 e^{-z/2}
\cosh (sz) \Bigrb) ,
\end{equation}
with $s= x_0/L-1/2$. Carrying out one integration, we find that
\begin{equation}
P(\gamma) \simeq 2 F \int_{z_1}^{z_2} \frac{dz}{z} \exp \left[
-\frac{(z-z_0)^2}{4 z_0} \right] \frac{1}{\sqrt{\gamma^2 - 4 e^{-z}}} ,
\label{pogamma}
\end{equation}
with $z_1=2\ln{(2/\gamma)}$, \ $z_2=+\infty$ if $0\!<\!\gamma\!<\!1$,
or $z_2=-\ln{(\gamma-1)}$ if $1\!<\!\gamma\!<\!2$.  We cannot
simplify Eq.~(\ref{pogamma}) further, and in Fig.~\ref{pgammanum} we
show the result of a numerical integration. Obviously, our estimate
of $P(\gamma)$ is accurate only when $\gamma \! \ll \!1$. Working the
various asymptotic limits, we see that for very small $\gamma$ ($z_1
\! \gg \! z_0$), the probability distribution vanishes as
\begin{equation}
\label{pogamma1}
P(\gamma) \propto \frac{z_0}{z_1^{5/4}} \ \exp{\left(
-\frac{z_1^2}{4z_0} \right)} .
\end{equation}
The distribution has a maximum around $\gamma_{mp} \simeq 2 e^{-z/2}$,
which becomes more pronounced for increasing disorder, namely,
\begin{equation}
\label{pogamma2}
P(\gamma_{mp}) \propto  \frac{1}{z_0} \ \exp{\left(\frac{z_0}{2}\right)} .
\end{equation}
The large values of $\gamma$ are not exponentially damped by disorder, as
we have
\begin{equation}
\label{pogamma3}
P(\gamma\simeq 1) \propto \frac{1}{z_0} .
\end{equation}
The knowledge of $P(\gamma)$ allows us to estimate various averages.
For instance, let us show that the model is consistent. From the
Breit-Wigner form of the conductance (\ref{gBW}) we can write
\begin{eqnarray}
\label{avloco}
\langle \ln{g} \rangle & = & \frac{1}{\Delta} \ \left\langle
\int_{-\Delta/2}^{\Delta/2} dE \ \ln \left[
\frac{\gamma^{(l)}\gamma^{(r)}}{(E/c\Delta)^2 + \gamma^2/4} \right]
\right\rangle \nonumber \\ & = & \left\langle \ln \left[
\gamma^{(l)}\gamma^{(r)} \right] \right\rangle -
\left\langle \ln \left(\frac{\gamma^2}{4}\right) \right\rangle -
\left\langle \ln \left(\frac{1}{\gamma^2 c^2} + 1 \right) \right\rangle
\nonumber \\ & & +
2 \left[ 1-\left\langle \gamma c  \ \arctan \left( \frac{1}{\gamma c}
\right) \right\rangle \right] ,
\end{eqnarray}
with the averages taken with respect to $P(\gamma)$. The first term on
the right-hand side gives the dominant contribution,
\begin{equation}
\label{avlglgr}
\left\langle \ln{ \left[ \gamma^{(l)}\gamma^{(r)} \right] } \right\rangle =
\langle -z \rangle \simeq -z_0 .
\end{equation}
Since $P(z)$ selects values of $z \! \sim \! z_0 \! \gg \! 1$, the
remaining terms of (\ref{avloco}) give the next leading-order
contribution, which is independent of $z_0$,
\begin{equation}
\label{avlgnot}
-\left\langle \ln \left[\left(\frac{\gamma}{2} \right)^2
\left(\frac{1}{\gamma^2 c^2} + 1 \right) \right] \right\rangle + 2
\ \simeq \ 2\ln(2c) + 2 .
\end{equation}

The fluctuations of $\ln{g}$ are characterized by the correlation function
\begin{equation}
\label{corrlogg}
{\cal C}_{g}(\Delta E) = \langle \ln g(E+\Delta E)\ln g(E) \rangle -
\langle \ln g(E) \rangle^2 ,
\end{equation}
with the average running over energy $E$ and disorder. Following the
statistical assumptions introduced above, we have
\begin{eqnarray}
\label{avlocoi2}
\langle \ln^2{g} \rangle  & = & \frac{1}{\Delta} \ \left\langle
\int_{-\Delta/2}^{\Delta/2} dE \ \ln^2 \left[
\frac{\gamma^{(l)}\gamma^{(r)}}{(E/c\Delta)^2 + \gamma^2/4} \right]
\right\rangle \nonumber  \\
& \simeq & \left\langle \ln^2 \left[ \gamma^{(l)} \gamma^{(r)}
\right] \right\rangle + \left\langle \ln^2 \left(
\frac{\gamma^2}{4} \right) \right\rangle + 2 \left\langle
\left\{ \ln \left[\gamma^{(l)} \gamma^{(r)} \right] - \ln \left(
\frac{\gamma}{2} \right) \right\} \left[ \ln \left(\gamma^2c^2 \right)
+ 1 \right] \right\rangle \nonumber  \\ & & +
\left\langle \ln^2 \left( \gamma^2 c^2 \right) \right\rangle
+ 4 \left\langle \ln \left( \gamma^2 c^2 \right) \right\rangle \nonumber \\
& \simeq & z_0^2 - 2 z_0 \left[ 2\ln (2c) + 1 \right] .
\end{eqnarray}
The variance of the distribution of $\ln{g}$ is then given by
${\cal C}_{g}(\Delta E\!=\!0) \simeq 2 z_0$, showing that our resonant
model for localized transport is consistent with the standard log-normal
distribution of the conductance \cite{Pichard} characterized by
Eq.~(\ref{LN}). In subsection~\ref{sec:localized}.C, we will apply the
resonant model to the correlation function (\ref{corrlogg}) and will
determine its energy-correlation length.

%------------------------------------------------------------------------
\subsection{Wigner time in the localized regime and correlations with the
conductance}

The resonant model can be applied to the Wigner time, whose
average is given by
\begin{eqnarray}
\label{avt}
\langle \tw \rangle & \simeq & \frac{\hbar}{2N} \ \frac{1}{c \Delta^2}
\left \langle \int_{-\Delta/2}^{\Delta/2} dE \ \left[
\frac{\gamma}{(E/c\Delta)^2 + \gamma^2/4} \right] \right\rangle
\nonumber \\ & \simeq & \frac{\hbar}{2N} \ \frac{4}{\Delta} \
\left\langle \arctan{\left(\frac{1}{\gamma c}\right)} \right\rangle
\nonumber \\ & \simeq & \frac{\hbar}{2N} \ \frac{2\pi}{\Delta} .
\end{eqnarray}
This result also checks the consistency of our model. Notice that
although $\tw$ fluctuates strongly with energy and from sample to
sample, $\langle \tw \rangle$ essentially does not depend on
disorder. This is due to the fact that the mean slope of the eigenphases
as a function of energy is proportional to the density of states
\cite{JP,DoSm} (or equal to zero, depending on the particular convention
adopted for $S$). Large values of $\tw$ at the resonances are compensated
by the flat parts in between. These large fluctuations are not
appropriately represented by the usual linear correlation function
(\ref{eq:taucor}), which, within our resonant model, is given by
\begin{equation}
C_\tau(\Delta E) = \left(\frac{\hbar}{2N}\right)^2 \ \frac{4 \pi}{c
\Delta^2} \ \int_{0}^{2} \ d \gamma \ \frac{P(\gamma)}{\gamma} \ \frac{1}
{[\Delta E/(c \gamma \Delta)]^2 +1} -
\langle \tw \rangle^2 .
\end{equation}
In order to see why, we calculate the second moment of $\tw$:
\begin{eqnarray}
\label{vartt}
\langle \tw^2 \rangle & = & \left(\frac{\hbar}{2N}\right)^2
\frac{8 \pi}{c \Delta^2} \ \int_{0}^{\infty} d z \ \frac{P(z)}{z} \
\int_{\gamma_{\rm min}}^{\gamma_{\rm max}} \frac{d \gamma}{\gamma
\sqrt{\gamma^2-\gamma_{\rm min}^2}} \nonumber \\
& \simeq & \left(\frac{\hbar}{2N}\right)^2 \ \frac{2 \pi^2}{c \Delta^2} \
\frac{1}{z_0} \ \exp \left( \frac{3}{4}z_0 \right) .
\end{eqnarray}
The lower and upper limits of $\gamma$ for a given $z$ are $\gamma_{\rm min}
= 2 e^{-z/2}$ and $\gamma_{\rm max}=1+e^{-z}$. In obtaining the above
result we have used the fact that the relevant values of $z$ are of the order
of $z_0 \! \gg \! 1$. The (exponentially) large fluctuations of $\tw$
reflect a very wide distribution. Hence, we will describe these fluctuations
in terms of the logarithm of $\tw$. Moreover, for the remaining of
this section we will adopt the dimensionless Wigner time
\begin{equation}
\label{dlesswg}
\tau=\frac{\tw}{\langle \tw \rangle} = 2 N \ \frac{\tw}{t_{\rm H}} ,
\end{equation}
where $t_{\rm H}=2\pi\hbar/\Delta$ is the Heisenberg time. Similarly to
the calculation of the average log-conductance, we write
\begin{eqnarray}
\label{avlotau}
\langle \ln{\tau} \rangle & = & \frac{1}{\Delta} \ \left\langle
\int_{-\Delta/2}^{\Delta/2} dE \ \ln^2 \left[
\frac{\gamma/(2\pi c)}{(E/c\Delta)^2 + \gamma^2/4} \right] \right\rangle
\nonumber \\ & \simeq & \langle \ln \gamma \rangle + 2 +
\ln{\left(\frac{4c}{\pi}\right)} .
\end{eqnarray}
The first term in the right-hand side gives the dominant contribution and,
to leading order in $z_0$, we have
\begin{equation}
\label{avlotau2}
\langle \ln{\tau} \rangle \simeq -\frac{z_0}{4} .
\end{equation}
Therefore, in the asymptotic limit of $z_0 \! \gg \! 1$,
the mean logarithmic of the conductance and the Wigner time
are simply proportional to each other,
\begin{equation}
\label{avlotau3}
\langle \ln{\tau} \rangle = \frac{1}{4} \ \langle \ln{g} \rangle .
\end{equation}
To investigate the fluctuations around this average value, we proceed
in a way analogous to Eq.~(\ref{avlocoi2}):
\begin{eqnarray}
\label{avlotaui2}
\langle \ln^2\tau \rangle & = & \frac{1}{\Delta} \ \left\langle
\int_{-\Delta/2}^{\Delta/2} dE \ \ln^2 \left[
\frac{\gamma/(2\pi c)}{(E/c\Delta)^2 + \gamma^2/4} \right] \right\rangle
\nonumber \\ & \simeq & \langle\ln^2 \gamma\rangle \simeq \frac{z_0^2}{12} .
\end{eqnarray}
The variance of $\ln \tau$ is then given by
\begin{equation}
\label{varlntau}
\mbox{var}(\ln{\tau}) \ = \ {\cal C}_{\tau}(\Delta E\!=\!0) \
\simeq \ \frac{z_0^2}{48} .
\end{equation}

The existence of correlations between the conductance and the Wigner
time, as we presented in the Introduction (Fig.\ref{preview}.b), can be
easily understood from Eqs.~(\ref{gBW})-(\ref{twBW}). Moreover, our
resonant transmission model for localized transport allows us to
quantify such correlations. For this purpose, instead of working with
the cross-correlator (\ref{eq:corrgt}), we define the logarithmic
correlator
\begin{equation}
{\cal C}_{g\tau}(\Delta E) = \frac{\langle \delta \ln g (E + \Delta E)
\ \delta \ln \tau (E) \rangle} {\langle (\delta \ln g)^2 \rangle^{1/2}
\langle (\delta \ln \tau )^2 \rangle^{1/2}}
\label{eq:corrgtilog}
\end{equation}
and calculate
\begin{eqnarray}
\label{avlocotau}
\langle \ln{g} \ln{\tau} \rangle & = & \frac{1}{\Delta} \ \left\langle
\int_{-\Delta/2}^{\Delta/2} dE \ \ln \left[
\frac{\gamma^{(l)}\gamma^{(r)}}{(E/c\Delta)^2 + \gamma^2/4} \right]
\ \ln \left[ \frac{\gamma/(2\pi c)}{(E/c\Delta)^2 + \gamma^2/4} \right]
\right\rangle \nonumber \\ & \simeq & \left\langle \ln \left[
\gamma^{(l)}\gamma^{(r)} \right] \ln \gamma  \right\rangle +
\left[\ln \left(\frac{2c}{\pi}\right)+2 + 2\ln(c) \right] \left\langle \ln
\left[\gamma^{(l)}\gamma^{(r)} \right] \right\rangle \nonumber \\ & & +
2[\ln(2c)+1]\langle \ln \gamma \rangle
\nonumber \\ & \simeq & \frac{z_0^2}{4} -  z_0
\left[\ln \left(\frac{2}{\pi}\right) + \frac{3}{2} \ln (2c) +
 2 \right] .
\end{eqnarray}
To leading order in $z_0$ we have
\begin{equation}
{\cal C}_{g\tau}(\Delta E\!=\!0) \simeq \sqrt{\frac{6}{z_0}} ,
\label{eq:corrgtilog2}
\end{equation}
showing that the cross-correlation decays slowly (not exponentially)
with disorder.

In Table I we present our simulations in the localized regime for
different disorder ($W$) and number of modes ($N$) in a stripe with
aspect ratio $L_x/L_y=4$. The results displayed show a qualitative
agreement with the predictions of our resonant model
[Eqs.~(\ref{avlglgr}), (\ref{LN}), (\ref{avt}), (\ref{avlotau2}),
(\ref{varlntau}), and (\ref{eq:corrgtilog2})]. For the range of
disorder that we are able to simulate, it is likely that the next-leading
order in the large-$z_0$ asymptotic expansions of the above equations is
required for a quantitative agreement. The $c-$dependent terms should
be calculated consistently with the value of $z_0$ that best describes
each sample. We will not attempt here this detailed comparison between
the model and our numerical simulations because we would need a much
better statistics than the one we dispose. The relationship (\ref{LN})
between the mean logarithmic conductance and its variance is only
approximately verified in our numerical simulations. It has already
been noticed \cite{AviPicMut} that in order to obtain the agreement
with Eq.~(\ref{LN}) one should simulate long stripes with weak disorder.

The fluctuations in $\ln{g}$ are larger than those in $\ln{\tau}$,
despite the fact that in our resonant model the former depends linearly
on $z_0$, while the latter goes quadratically with $z_0$. It is the
difference in the prefactors that makes, for the values of the
disorder that we have simulated, the fluctuations of $\ln{g}$ larger.
One interesting aspect of the last column of Table I, the value of the
cross correlations, is that the results of simulations seem to show an
even slower dependence on $z_0$ than that of Eq.~(\ref{eq:corrgtilog2}).

%------------------------------------------------------------------------
\subsection{Energy and parametric correlations}

\begin{figure}
\setlength{\unitlength}{1mm}
\begin{picture}(200,100)
\put(0,0){\epsfxsize=13cm\epsfbox{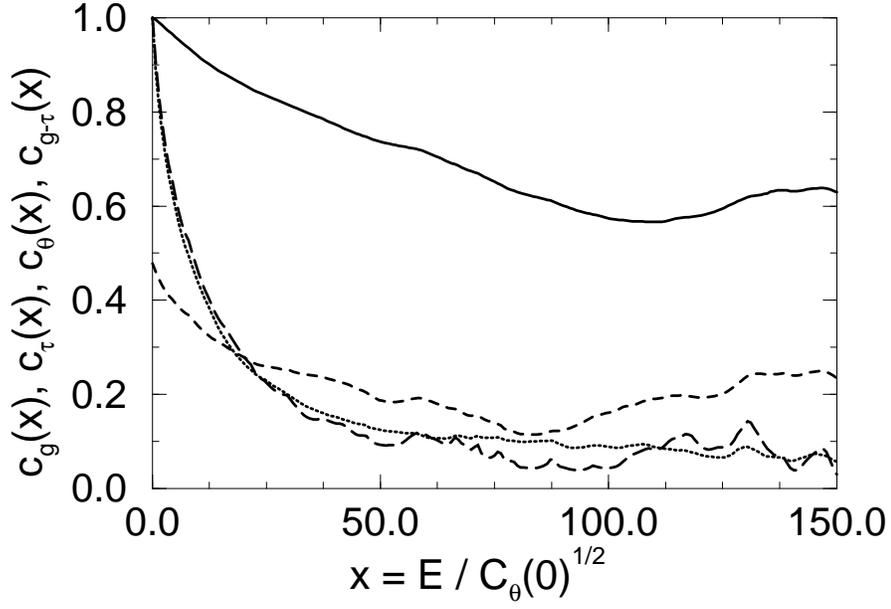}}
\end{picture}
\caption{Logarithmic autocorrelations of the conductance (solid),
Wigner time (dotted), and eigenphase velocity (long-dashed) for the
strongly localized case ($W$=4 in Anderson units) without magnetic
field. The cross-correlation between $g$ and $\tw$ is shown by the
thick short-dashed curve. The use of the logarithmic was required
in view of the broad distributions of $g$ and $\tw$ (see text).}
\label{loccorr}
\end{figure}

The energy-dependent correlation functions for the log-conductance and
log-Wigner time can be calculated from the resonant model along the same
lines that we followed in the previous subsections. In particular,
the various energy correlation lengths can be estimated from the initial
curvature of the corresponding correlation functions. For the energy
correlator of Eq.~(\ref{corrlogg}) we have
\begin{eqnarray}
\label{curvloco}
{\cal C}_{g}^{\prime\prime} (\Delta E\!=\!0) & \simeq &
\frac{4}{\Delta^2} \ \left\langle \frac{1}{\gamma c} \
\int_{-1/\gamma c}^{1/\gamma c}
dx \ \frac{x^2-1}{(x^2+1)^2} \ \left\{ \ln \left[
\gamma^{(l)}\gamma^{(r)} \right] - \ln \left( \frac{\gamma^2}{4}\right)
-\ln \left(x^2+1\right) \right\} \right\rangle \nonumber \\ &
\simeq & - \frac{8}{\Delta^2} \left\{ \frac{\pi}{2c} \left\langle
\frac{1} {\gamma} \right\rangle + \left\langle \ln \left[
\gamma^{(l)}\gamma^{(r)} \right] \right\rangle + 2 \ln (2c) -1 \right\} .
\end{eqnarray}
The term proportional to $\langle 1/\gamma \rangle$ gives the dominant
contribution and its calculation is analogous to that of Eq.~(\ref{vartt}),
yielding
\begin{equation}
{\cal C}_{g}^{\prime\prime} (\Delta E\!=\!0) \simeq
- \frac{2 \pi^2}{c\Delta^2} \ \frac{1}{z_0} \ \exp{\left(
\frac{3}{4}z_0 \right)} .
\end{equation}
Assuming that the correlation function ${\cal C}_{g}(\Delta E)$ has
approximately a Lorentzian form, its correlation energy will be related
to the curvature at the origin through
\begin{equation}
\Delta E_{g}^{c} \simeq \sqrt{-\frac{2{\cal C}_{g}(0)}
{{\cal C}_{g}^{\prime\prime}(0)}} \simeq \Delta
\frac{\sqrt{2c}}{\pi} \ z_0 \ \exp{\left(-\frac{3}{8}z_0\right)} .
\end{equation}
For the logarithmic correlation of the Wigner time we have (to leading
order in $z_0$) the same curvature than for ${\cal C}_{g}(\Delta E)$ since
$\ln{\tau}$ and $\ln{g}$ have the same energy dependence [Eqs. (\ref{gBW})-
(\ref{twBW})]. The difference in the correlation lengths $\Delta E_{g}^{c}$
and $\Delta E_{\tau}^{c}$ comes from the different values of the
variances ${\cal C}_{g}(0)$ and ${\cal C}_{\tau}(0)$, yielding
\begin{equation}
\Delta E_{\tau}^{c} \simeq \Delta \frac{1}{4\pi}
\sqrt{\frac{c}{3}} \ z_0^{3/2} \ \exp{\left( -\frac{3}{8}z_0\right)} .
\end{equation}
Although the resonant model predicts that ${\cal C}_{\tau}(0)$ and
${\cal C}_{g}(0)$ have a quadratic and a linear dependence on $z_0$,
respectively, we find numerically that $\mbox{var}(\ln{g})$ is always
larger than $\mbox{var}(\ln{\tau})$ within the range of disorder and
stripe lengths simulated. Therefore, we expect that $\Delta E_{g}^{c}
> \Delta E_{\tau}^{c}$, which is indeed the behavior observed in
Fig.~\ref{loccorr}. The energy-correlation length of ${\cal C}_{g\tau}$
is intermediate between $\Delta E_{g}^{c}$ and $\Delta E_{\tau}^{c}$.
In all three cases, the correlation length shrinks exponentially with
$z_0$, due to the fact that the conductance and Wigner-time peaks become
narrower with increasing disorder. This is the reason why we have used
in Fig.~\ref{loccorr} the energy rescaling appropriate for the parametric
correlations of eigenphase velocities which also turns out to depend
exponentially on $z_0$ (see below).

The correlation functions in the localized regime can be calculated
with the aid of the resonant transmission model. The analysis is more
involved for eigenphase velocities than for $g$ or $\tw$ because we do
not have a simple expression like (\ref{gBW}) and (\ref{twBW}) relating
eigenphases to energy. However, there are obvious relations between
correlators of eigenphases and Wigner times. The eigenphase velocity
introduced in Eq.~(\ref{eq:corphase}) is simply the diagonal part ($l=m$)
of the sum defining the Wigner time correlator of Eq.~(\ref{eq:taucor}).
In particular, $C_\theta(\Delta E)$ and $C_{\tau}(\Delta E)$ are identical
in one trivial limit of the localized case, namely, the $N=1$, when we
know the relation exactly (see Appendix \ref{appB}): Close to a resonance
with energy $E_{\nu}$, we have
\begin{equation}
\theta_{\pm}(E) \approx \pm \frac{\pi}{2} + \arctan [2(E-E_{\nu})/\Gamma],
\label{eq:etawm}
\end{equation}
from which we recover the Breit-Wigner form of (\ref{twBW})
\begin{equation}
\label{twBW1}
\tw(E) = \frac{\hbar}{2} \ \frac{\Gamma_{\nu}}{(E-E_{\nu})^2
+ \Gamma_{\nu}^{2}/4} .
\end{equation}
In the large-$N$ limit and for strongly disordered samples we can assume
that the $\{\theta_l\}$ move almost rigidly, repelling each other
simultaneously when a resonance occurs. Therefore, there is approximately
no distinction between diagonal and off-diagonal correlations because all
eigenphases follow a similar pattern of energy evolution This is confirmed
in our numerical simulations by the close agreement between
${\cal C}_\tau$ and ${\cal C}_\theta$ over a wide energy range. The large
fluctuations for the Wigner time translate into very large values of
$C_\theta(0)$, as evident from Fig.~\ref{preview}.b. (Notice that the sharp 
steps of the phase shifts give rise to very large derivatives.) The broad
distribution of eigenphases forces us to work with the logarithmic
correlator
\begin{equation}
{\cal C}_\theta(\Delta E) = \left\langle \ln \left[\frac{d\theta_l(E+\Delta
E)}{dE}\right] \ \ln \left[ \frac{d\theta_l(E)}{dE}\right]
\right\rangle - \left\langle \ln \left[ \frac{d\theta_l(E)}{dE}
\right] \right\rangle^2 ,
\label{eq:corphasel}
\end{equation}
but still adopting the rescaling used in the metallic regime
[Eq.~(\ref{eq:rescaltheta})],
\begin{mathletters}
\label{eq:rescalthetal}
\begin{eqnarray}
x & = & \Delta E\ \sqrt{{\cal C}_\theta(0)}\\ c_\theta(x) & = &
{\cal C}_\theta(\Delta E)/{\cal C}_\theta(0) .
\end{eqnarray}
\end{mathletters}
\nin The average in Eq.~(\ref{eq:corphasel}) is over energy $E$, disorder,
and channel index $l$. In Fig.~\ref{loccorr} we present $c_\theta(x)$ for
in the localized regime (disorder $W\!=\!4$). The universality found for the
corresponding eigenphase velocity of the metallic regime is lost in a way
consistent with our resonant model. On the other hand, when properly rescaled,
we obtain an agreement between $c_\theta(x)$ and the Wigner time correlator.

%%%%%%%%%%%%%%%%%%%%%%%%%%%%%%%%%%%%%%%%%%%%%%%%%%%%%%%%%%%%%%%%%%%%%%%%%
\section{Conclusions}
\label{sec:concl}

In this work we have studied parametric correlation functions in disordered
quasi-one dimensional systems in the metallic and localized regimes.
For this purpose we have considered the fluctuations in energy of the
eigenphases of the scattering matrix, as well as fluctuations of the
conductance and the Wigner time, and their cross correlation.

In the metallic regime, when disorder is weak and the fluctuations of
the eigenphases $\{\theta_l\}$ are well described by Dyson's circular
ensembles \cite{JP}, the parametric correlations obtained from our
numerical simulations follow closely the universal behavior discovered
by Szafer, Altshuler, and Simons \cite{Aaron,Ben} for the spectra of
chaotic and disordered systems. This finding is justified from a
Brownian-motion model similar to that developed by Beenakker \cite{Been3}
for the energy spectrum. The Brownian-motion model also allows us to
obtain the energy correlation function of the Wigner time in the
large-energy asymptotic limit. We have compared these analytical results
with our numerical simulations. Our simulations in the metallic regime
display a strong correlation between the conductance and the Wigner time,
which arises from the symmetry restrictions to the energy drift
of the scattering matrix.

Transport in the localized regime is resonant-like and, therefore,
its statistical properties are given by those of localized wave
functions in the disordered stripe. From the well-known fact that
wave functions are (1) localized around centers uniformly distributed
in the sample (2) with inverse localization lengths following a
Gaussian distribution, we recover the basic features observed in
the simulations where disorder was strong. Our model is consistent
with the standard log-normal distribution for the conductance and
allowed us to estimate the typical energy correlation length of
conductance fluctuations. The very large fluctuations of the Wigner
time led us to study the distribution of its logarithm; the
variance of this distribution was found to be related to that of
the conductance. We also investigated the correlation between the
conductance and the Wigner time as a function of disorder. The
energy-dependent parametric correlations in the localized regime
follow closely the behavior of the Wigner time.

An experimental check of some of the ideas developed in this
work has become possible with the fabrication of very short
insulating wires \cite{Lad,Jer}. In particular, the Breit-Wigner
line shape characteristic of resonant tunneling has been established
and the statistics of peak position (as a function of Fermi energy
or gate voltage) has been shown to display level repulsion \cite{Jer}.
This peculiar finding can be been understood within a resonant-tunneling
picture if the observed peaks correspond to states well connected
to the leads, which are confined to a small region in center of
the stripe and which are therefore not necessarily separated from
each other by distances much larger than the localization length.
But the experimental study of short wires poses also the question of
whether the single-particle approach that we have pursued in this
work is valid. At least in a strictly one-dimensional sample it is
known \cite{LuttLiqu} that two-body potentials drive the
system to a non-Fermi liquid behavior, leading to a more sensitive
dependence on weak disorder \cite{Kane92}. Moreover, given the recent
developments stressing the interplay between disorder and interactions
in the localized regime \cite{Shep,Wei,vonOpp}, it would be interesting
to extend the present study in order to incorporate the effects of
electron-electron interactions in the energy-dependent parametric
correlations.

%%%%%%%%%%%%%%%%%%%%%%%%%%%%%%%%%%%%%%%%%%%%%%%%%%%%%%%%%%%%%%%%%%%%%%%%%
\acknowledgements

We are grateful to Boris Altshuler, who participated in the early stages
of this work and provided us with many illuminating comments. We also
acknowledge helpful discussions with C. Beenakker, P. Brouwer, C.
Lewenkopf and P. A. Mello. This research was supported in part by the
National Science Foundation under Grant No. PHY94-07194. E.R.M. and
R.A.J thank the kind hospitality and support of NORDITA, where this work
was initiated.

%%%%%%%%%%%%%%%%%%%%%%%%%%%%%%%%%%%%%%%%%%%%%%%%%%%%%%%%%%%%%%%%%%%%%%%%%
\appendix
\section{Statistical properties of the Wigner-Smith matrix}
\label{appA}

 In this appendix we study the statistical distribution of the
Wigner-Smith matrix in order to investigate more carefully the
applicability of a unitary Brownian-motion model to the energy
dependence of the scattering matrix.

For a small energy displacement, the evolution of the scattering matrix
can be written as
\begin{equation}
S(E+\delta E) = S(E) \ \exp{(i\delta K)} ,
\label{brown}
\end{equation}
where $\delta K$ is required to be an infinitesimal Hermitian matrix
to preserve the unitarity of $S$.
To first order in $\delta E$, we can write
\begin{equation}
\delta K \ = \ -i \ S^{\dagger} \frac{dS(E)}{dE} \ \delta E \ =
\ \frac{2N}{\hbar} \ Q \ \delta E .
\label{relXQ}
\end{equation}
Consequently, considering now the form (\ref{eq:dS}) of the allowed
infinitesimal variations in the neighborhood of $S$, we can identify
\begin{equation}
Q \ \delta E = \frac{\hbar}{2N} \ Y^{\dagger} \ (\delta \HT) \ Y ,
\label{eiQ}
\end{equation}
showing that the eigenvalues of $Q$ and $\delta \HT$ are proportional to
each other. Moreover, if $S$ has an isotropic distribution at a given
$E$ and remains isotropically distributed for each energy in the interval
$\delta E$, one can see that this property should characterize $Q$ as 
well. The isotropy of $S$ and $Q$ are related. We now try to determine 
which are the requirements on the probability distribution of $Q$ necessary
for the validity of the BMM [Eqs.~(\ref{alleq:momQs})] at least in an 
approximate sense and for a sufficiently small $\delta E$.

\begin{figure}
\setlength{\unitlength}{1mm}
\begin{picture}(200,150)
\put(0,0){\epsfxsize=13cm\epsfbox{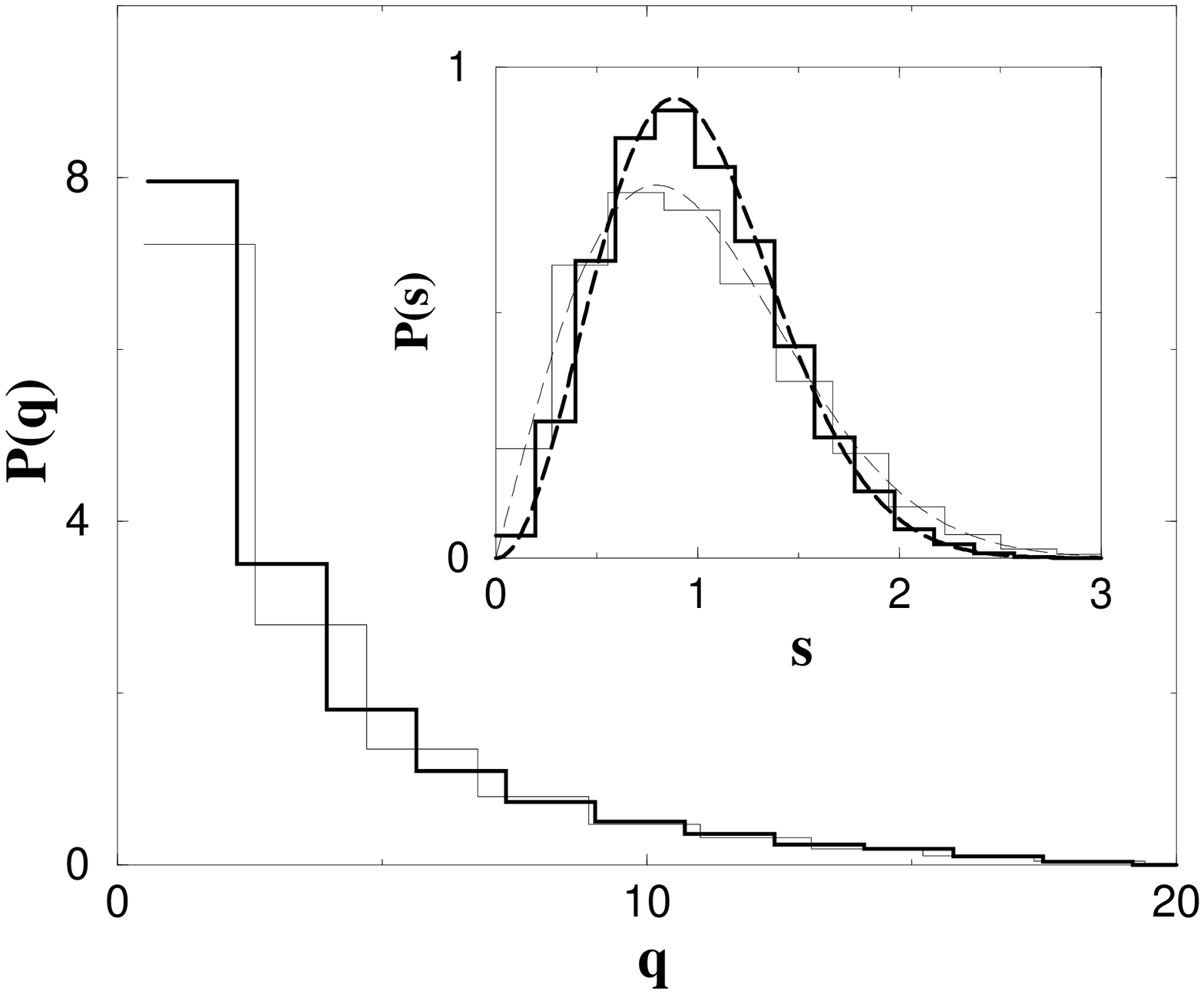}}
\end{picture}
\caption{Distribution of the eigenvalues of the Wigner-Smith matrix
in the weakly-disordered metallic regime. The thick and thin lines
correspond to the presence or absence of a magnetic field, respectively.
The inset shows the nearest-neighbor spacing histogram for the same
eigenvalues (similar convention for the line widths). The Wigner
surmises for the GUE and GOE (thick and thin dashed lines, respectively)
are shown for comparison.}
\label{WigSmi}
\end{figure}

We have investigated the probability distribution of the eigenvalues
of $Q$ by a numerical simulation on weakly-disordered metallic stripes.
Under these conditions, the correlations of the eigenphases of $S$ are
well approximated by those of the CUE (or COE) ensemble \cite{JP}, and
the eigenphase velocity correlation agrees well with those found for
chaotic Hamiltonian systems. In Fig.~\ref{WigSmi} we present the mean
density of eigenvalues of $Q$ and its nearest-neighbor spacing distribution
(inset) for the cases without and with a time-symmetry breaking
magnetic field. Notice that, within our convention, all eigenvalues of
$Q$ are positive, consistently with their interpretation as typical
traversal times through the disorder region. Due to this symmetry
requirement (that rules out the usual Gaussian ensembles), one of the
simplest possible random-matrix description of $Q$ is provided
by the Laguerre ensembles. The observed eigenvalue density and the
nearest-neighbor distribution (after unfolding \cite{BGS}) are not
incompatible with such an ensemble at first sight. Indeed, one can 
see a clear level repulsion and a good agreement with the
Wigner's surmise of the appropriate symmetry class. Notice that although 
$Q$ is not symmetric, the nearest-neighbor distribution in the 
time-reversal-symmetric case is given by the $\beta\!=\!1$
Wigner surmise. This is because the submanifold of allowed $Q$ matrices
in the presence of time-reversal symmetry can be mapped into
that of the real symmetric ones by a transformation that leaves the
eigenvalues unchanged \cite{MePri}. It is important to remember that the
other set of characteristic times, the energy derivatives of the phase
shifts $\{d\theta_l/dE\}$ associated with the eigenchannels of $S$,
{\it do not} exhibit nearest-neighbor repulsion. Both sets are, however,
obviously related since their sum is simply $\tw$.

For a disordered wire, $S$ is not isotropic. (However, the circular 
ensembles give a good description of the phase shift fluctuations in 
the metallic regime.) Since the isotropy of $S$ and $Q$ are related, 
$Q$ cannot be isotropic for a disordered wire. But one can hope that
this does not matter for the spectral fluctuations of $Q$, in analogy
with what happens for the spectral fluctuations of $S$. This lead us to
consider a distribution for the Wigner-Smith matrix that is invariant
under unitary (orthogonal) transformations, and to propose a simplified
maximum-entropy ansatz \cite{Balian,St}; in other words, we adopt a 
maximum-entropy distribution for $Q$, given the observed mean density of 
eigenvalues. This will yield a logarithmic interaction between eigenvalues 
and consequently the observed level repulsion. In the usual Coulomb gas
analogy, one would have a certain (non-parabolic) confining
potential for this ansatz giving the observed mean density of positive
eigenvalues. The problem we face is then quite analogous to that of
the probability distribution of the radial parameters of the transfer
matrix \cite{St,Been2}.

It is well known that a maximum-entropy approach with an arbitrary confining
potential leads to correlations between matrix elements \cite{Porter}. The
eigenvector isotropy assumption allows us to average over the unitary
group \cite{Mello,Mehtabook}, yielding
\begin{mathletters}
\label{allqmes}
\begin{equation}
\langle Q_{ij} \rangle = \delta_{ij} \ \frac{\langle \mbox{Tr} Q \rangle}{N}
\label{eq1}
\end{equation}
and
\begin{equation}
\langle Q_{ij} Q_{kl} \rangle = \frac{1}{N^2-1} \left[ \delta_{ij}\delta_{kl}
\left(\sum_{mn} \langle q_m q_n \rangle - \frac{1}{N} \sum_{n} \langle q_n^{2}
\rangle
\right) + \delta_{il}\delta_{jk} \left(\sum_{n} \langle q_n^{2} \rangle -
\frac{1}{N} \sum_{mn} \langle q_m q_n \rangle \right) \right]
\label{eq2}
\end{equation}
\end{mathletters}
Above, brackets on the left-hand sides imply averages over the probability
distribution of $Q$, while brackets on the right-hand sides indicate
averages over the eigenvalue distribution. Equation (\ref{eq1}) can be
trivially set to zero once we suppress the constant drift of the eigenphases
with energy (or use the convention $\langle \tw \rangle = 0$, like in
subsection~\ref{sec:metallic}.c). The non-vanishing correlations of
Eq.~(\ref{eq2}) are
\begin{mathletters}
\label{allqnvs}
\begin{equation}
\langle Q_{ii}^2 \rangle = \frac{1}{N+1} \left[ (N-1) \langle q_1q_2
\rangle +2 \langle q_1^2\rangle \right] ,
\label{eqnv1}
\end{equation}
\begin{equation}
\langle Q_{ii} Q_{jj} \rangle = \frac{1}{N+1} \left[ N \langle q_1q_2
\rangle + \langle q_1^2\rangle \right] \hspace{1cm} i \neq j ,
\label{eqnv2}
\end{equation}
\begin{equation}
\langle Q_{ij} Q_{ji} \rangle = \frac{1}{N+1} \left[ \langle q_1^2
\rangle - \langle q_1q_2 \rangle \right] \hspace{1cm} i \neq j .
\label{eqnv3}
\end{equation}
\end{mathletters}
It is only in the case of the quadratic confinement potential that we
have the GUE result $\langle q_1q_2 \rangle=-\langle q_1^2\rangle /N$
implying a vanishing correlation between different diagonal elements
[Eq.~(\ref{eqnv2})] and the usual correlation ($\langle Q_{ij} Q_{ji}
\rangle = \langle q_1^2\rangle /N$) between symmetric matrix elements
[Eqs.~(\ref{eqnv1}) and (\ref{eqnv3})]. Other confining potentials,
like the one compatible with the observed mean density, result
in different two-point correlation functions $\langle q_1q_2 \rangle$
and nonvanishing matrix element correlations in all Eqs.~(\ref{allqnvs}).

If we assume that the eigenvectors of $\delta\HT$ are also isotropically
distributed, Eqs.~(\ref{eiQ}) and (\ref{allqnvs}) show that there exist
correlations between different matrix elements of $\delta\HT$. Therefore,
Eqs.~(\ref{alleq:momQs}) cannot be verified exactly and their validity is
actually approximate. On the other hand, the correlations of the type of
Eq.~(\ref{eqnv2}) decrease with $N$ and in the large $N$ limit the
maximum-entropy ensembles provide a local approximation to the Gaussian
ones \cite{ozor:book,BGS}. Therefore, it is only in the large $N$ limit
that the BMM for $S$ can be appropriate.

Assuming a nearly Gaussian behavior and comparing Eqs.~(\ref{eqnv1})
and (\ref{eqnv3}) with Eq.~(\ref{alleq:momQs}) allows the identifications
\begin{equation}
\delta t = (\delta E)^2
\label{ident1}
\end{equation}
and
\begin{equation}
\frac{1}{f}= \frac{2 \beta}{\hbar^2} \ \langle \mbox{Tr} Q^2 \rangle ,
\label{ident2}
\end{equation}
of the fictitious time and the friction coefficient of the BMM
with the energy and Wigner-Smith matrix associated with the
scattering process.

%%%%%%%%%%%%%%%%%%%%%%%%%%%%%%%%%%%%%%%%%%%%%%%%%%%%%%%%%%%%%%%%%%%%%%%%
\section{Scattering in the one-channel case}
\label{appB}

In this appendix we calculate the energy dependence of the eigenphases
around a resonance for the one channel case.
The starting point is the S-matrix expression \cite{Mahaux69}
\begin{equation}
S_{ab}(E) = e^{2i\varphi_a(E)} \delta_{ab} - 2\pi i\ e^{i\varphi_a(E) +
i\varphi_b(E)} \sum_{\nu\mu} W^\ast_{a\mu}(E) [D^{-1}(E)]_{\mu\nu}
W_{b\nu}(E) ,
\label{eq:SmatWM}
\end{equation}
with
\begin{equation}
D_{\mu\nu}(E) = E\delta_{\mu\nu} - H_{\mu\nu} + i\pi \sum_c
W^\ast_{c\mu}(E) W_{c\nu}(E) .
\label{eq:discrim}
\end{equation}
$W_{a\mu}(E)$ represents the overlap between the external
wave functions (plane waves) and the internal eigenfunctions of $H$
(for a time-reversal symmetric system we can choose $W_{a\mu}$ to be
real). The usual approximation is to assume that the energy dependence
of the phases $\varphi_a(E)$ and matrix elements $W_{a\mu}(E)$ is
smooth over the interval where there are many resonances (poles) in
$D^{-1}(E)$. Moreover, if $|W_{a\mu}|^2$ is typically much smaller than
the average distance between poles, we can expand $[D^{-1}(E)]$ and perform
a unitary transformation in the wave functions to obtain
\begin{equation}
S_{ab}(E) = \delta_{ab} - 2\pi i \sum_{\nu} \frac{W^\ast_{a\nu} W_{b\nu}}
{E - E_\nu + i\Gamma_\nu/2} + O(\Gamma/\Delta) ,
\label{eq:Smatres}
\end{equation}
where $\Gamma_\nu = 2\pi\sum_c |W_{c\nu}|^2$. Now, specializing for a
one-dimensional system (therefore $N=1$) and looking at energies close to
a certain resonance, we can get explicit expressions for the coefficients
\begin{mathletters}
\label{eq:1DSelem}
\begin{eqnarray}
r & \approx & 1 - \frac{2\Gamma_\nu^{(R)}/\Gamma_\nu}
{1 - 2i(E-E_\nu)/\Gamma_\nu}  \\
r^\prime & \approx & 1 - \frac{2\Gamma_\nu^{(L)}/\Gamma_\nu}
{1 - 2i(E-E_\nu)/\Gamma_\nu} \\
t & \approx & -\frac{\alpha}{1 - 2i (E-E_\nu)/ \Gamma_\nu} \\
t^\prime & \approx & -\frac{\alpha^\ast}{1 - 2i (E-E_\nu)/\Gamma_\nu} ,
\end{eqnarray}
\end{mathletters}
with $\alpha = 4\pi W_{L\nu}^\ast W_{R\nu}/\Gamma_\nu$. Notice that
(\ref{eq:Smatres}) keeps $S$ unitary only to lowest order in $\Gamma/\Delta$.

It is useful to introduce the following parameterization for $S$:
\begin{eqnarray}
r & = & \sqrt{R}\ e^{i\eta + \chi} \nonumber \\
r^\ast & = & \sqrt{R}\ e^{\eta -\chi} \nonumber \\
t & = & i\sqrt{T}\ e^{i\eta + i\kappa} \nonumber \\
t^\ast & = & i\sqrt{T}\ e^{i\eta - i\kappa},
\label{eq:paramet}
\end{eqnarray}
with $R+T=1$. The various quantities appearing above can be determined
through Eq.~(\ref{eq:1DSelem}); in particular,
\begin{equation}
\eta(E) \approx \arctan [2(E-E_\nu)/\Gamma_\nu]
\label{eq:eta}
\end{equation}
and
\begin{equation}
T(E) \approx \frac{|\alpha|^2}{1 + 4(E-E_\nu)^2/\Gamma_\nu}.
\label{eq:Transm}
\end{equation}
For time-reversal symmetric systems ($\kappa=0$), it is easy to find
an expression for the eigenphases of $S$ in terms of the new parameters,
namely,
\begin{equation}
\theta_\pm = \eta \pm \arctan \left( \sqrt{\frac{T}{R}} \right), \ \ \
\mbox{mod}(\pi) .
\end{equation}
Using Eqs.~(\ref{eq:eta}) and (\ref{eq:Transm}), we then have
\begin{equation}
\theta_\pm(E) \approx \arctan [2(E-E_\nu)/\Gamma_\nu] \pm \arctan
\left[ \frac{|\alpha|^2}{ 1- |\alpha|^2 + 4(E-E_\nu)^2/
\Gamma_\nu} \right]^{1/2}, \ \ \ \mbox{mod}(\pi) .
\label{eq:phaseappro}
\end{equation}
Notice that, in general, $\Gamma_R\ne\Gamma_L$ and $|\alpha|^2< 1$. This
means that the second term on the r.h.s of Eq.~(\ref{eq:phaseappro})
varies slower than the first and the difference between eigenphases 
around a resonance is approximately $\pi$ [Eq.~(\ref{eq:etawm})].

\vspace{3ex}

%%%%%%%%%%%%%%%%%%%%%%%%%%%%%%%%%%%%%%%%%%%%%%%%%%%%%%%%%%%%%%%%%%%%%%%%

\nin {\em Note added:} While finishing this manuscript, we learnt
from C. Beenakker that an exact random matrix description
of the distribution of the Wigner-Smith matrix $Q$ can be obtained
in the case that either $S$ remains distributed following the
Dyson circular ensembles as $E$ varies, or the underlying
Hamiltonian is a member of the Gaussian ensembles.
This can be applied to ballistic chaotic cavities, but not to the
disordered wires which we study. However, the result is very
similar to the maximum-entropy description which we proposed in
the Appendix {\ref{appA}}, notably as far the level repulsion is
concerned. The extension of the derivation made by
Beenakker and co-workers from the ballistic cavity to the disordered
wire should allow us to see what corrections to the 
maximum-entropy ansatz proposed in Appendix \ref{appA} are required.

%%%%%%%%%%%%%%%%%%%%%%%%%%%%%%%%%%%%%%%%%%%%%%%%%%%%%%%%%%%%%%%%%%%%%%%%

%%%%%%%%%%%%%%%%%%%%%%%%%%%%%%%%%%%%%%%%%%%%%%%%%%%%%%%%%%%%%%%%%%%%%%

\begin{table}
\begin{tabular}{|cccccccc|}
\ W \         & \ $N$ \ & \ $\langle \ln{g} \rangle \ $ \  & \
	$\langle (\delta \ln{g})^2 \rangle$ \
	& \ $\langle \tau \rangle \ $ \   & \ $\langle \ln{\tau} \rangle
          \ $ \
        & \ $\langle (\delta \ln{\tau})^2 \rangle$ \
	& ${\cal C}_{g\tau}$ \ \\ \hline\hline
4  &14 &-4.1  &7.0  &1.3 &-0.01  &0.40 &0.67   \\
4  &10 &-6.2  &9.2  &1.2 &-0.21  &0.62 &0.74   \\
5  &14 &-9.8  &8.6  &1.2 &-0.43  &0.71 &0.62   \\
5  &10 &-9.0  &13.8 &1.1 &-0.49  &0.79 &0.60   \\
6  &14 &-16.1 &29.1 &1.2 &-0.69  &0.80 &0.53   \\
6  &10 &-17.2 &28.8 &1.3 &-0.99  &0.93 &0.55   \\
7  &14 &-24.5 &39.3 &1.1 &-0.92  &0.81 &0.52   \\
\end{tabular}

\vspace{2ex}

\caption{
Simulations of stripes in the localized regime with aspect ratio
$L_x/L_y=4$, mean disorder $W$, and number of channels $N$. The
averages are over energy and impurity configuration.
$\langle \ln{g} \rangle $ and $\langle (\delta \ln{g})^2 \rangle$
are the mean logarithmic conductance and its variance, respectively.
The mean Wigner time $\langle \tau \rangle \ $ has been normalized
to its average value $\pi\hbar/N \Delta$, and therefore the fifth
column checks the validity of Eq.~(\protect\ref{avt}).
$\langle \ln{\tau} \rangle$ and $\langle (\delta \ln{\tau})^2 \rangle$
are the mean logarithmic Wigner time and its variance, respectively.
The last column is the value of the cross correlator between the
conductance and Wigner time calculated as in
Eq.~(\protect\ref{eq:corrgtilog}).
}
\end{table}

%%%%%%%%%%%%%%%%%%%%%%%%%%%%%%%%%%%%%%%%%%%%%%%%%%%%%%%%%%%%%%%%%%%%%%

\newpage

\newpage

\newpage

\newpage

\newpage

\begin{references}
\bibitem{curves} The COE and CUE theoretical curves were obtained
according to Eqs.~(\ref{eq:corenergy}) after numerical diagonalization of
500$\times$500 matrices of the form $H(X) = H_1\cos(X) + H_2\sin(X)$,
where $H_1$ and $H_2$ were drawn from the proper Gaussian
ensemble. The averages were performed over the 100 central eigenvalues
and 50 realizations.
\end{references}
\end{document}